\documentclass[12pt,preprint]{aastex}

\shorttitle{Direct imaging search for $\epsilon$ Eri b}
\shortauthors{Janson et al.}

\begin{document}


\title{NACO-SDI direct imaging search for the exoplanet $\epsilon$ Eri b \footnote{Based on observations collected at the European Southern Observatory, Chile (NACO SDI commissioning run August 2003, ESO No.\ 273.C-5030, and ESO No.\ 076.C-0027)}}


\author{Markus Janson\altaffilmark{1}, Wolfgang Brandner\altaffilmark{1,9}, Thomas Henning\altaffilmark{1}, and Rainer Lenzen\altaffilmark{1}}
\email{janson@mpia.de}

\author{Barbara McArthur\altaffilmark{2} and G. Fritz Benedict\altaffilmark{2}}

\author{Sabine Reffert\altaffilmark{3}}

\author{Eric Nielsen\altaffilmark{4}, Laird Close\altaffilmark{4}, and Beth Biller\altaffilmark{4}}

\author{Stephan Kellner\altaffilmark{5}}

\author{Eike G\"unther\altaffilmark{6} and Artie Hatzes\altaffilmark{6}}

\author{Elena Masciadri\altaffilmark{7}}

\and

\author{Kerstin Gei{\ss}ler\altaffilmark{8} and Markus Hartung\altaffilmark{8}}


\altaffiltext{1}{Max-Planck-Institut f\"ur Astronomie, K\"onigstuhl 17, D-69117 Heidelberg, Germany}
\altaffiltext{2}{University of Texas, McDonald Observatory, 1 University Station, C1402, Austin, Texas 78712-0259}
\altaffiltext{3}{Landessternwarte, K\"onigstuhl 17, D-69117 Heidelberg, Germany}
\altaffiltext{4}{University of Arizona, 933 N. Cherry Ave, Tucson, AZ 85721}
\altaffiltext{5}{Keck Observatory, 65-1120 Mamalahoa Hwy, Kamuela, Hawaii 96743}
\altaffiltext{6}{Th\"uringer Landessternwarte, Sternwarte 5, D-07778 Tautenburg, Germany}
\altaffiltext{7}{Osservatorio Astrofisico di Arcetri, INAF, Largo Enrico Fermi 5, I-50125 Florence, Italy}
\altaffiltext{8}{European Southern Observatory, Alonso de Cordova 3107, Vitacura, Casilla 19001, Santiago 19, Chile}
\altaffiltext{9}{UCLA, Div. of Astronomy, 475 Portola Plaza, Los Angeles, CA-90095-1547, USA}



\begin{abstract}
The active K2V star $\epsilon$ Eri hosts the most nearby known extrasolar planet. With an angular separation of about 1'' on average, and an age of a few to several hundred Myrs, $\epsilon$ Eri b is one of the prime candidates for becoming the first definitive extrasolar planet imaged directly. We present a multi-epoch deep differential imaging survey performed with NACO-SDI at the VLT with the aim of finding the planet. The results are combined with recent astrometry in an attempt to further constrain the detection limits. No convincing candidate is found among the many coherent structures that constitute the residual speckle noise, which is the dominant noise at small angular scales. We present our detection limits, compare them with the estimated brightness of $\epsilon$ Eri b, and analyze how the limits can be improved further. It is found that integration time remains a very important parameter for achieving good results, even in the speckle-dominated regimes. The results yield new, improved upper 3$\sigma$ limits on the absolute H-band (1.6 $\mu$m) brightness of the 1.55 $M_{\rm jup}$ companion of 19.1 to 19.5 mag, depending on the specific age of the system.
\end{abstract}





\section{Introduction}

\label{sec_intro}

High-contrast imaging from the ground is a rapidly progressing field of astronomy. Developments of adaptive optics (AO) along with employment of innovative differential imaging techniques have led to a continuous improvement in terms of higher reachable contrasts at smaller separations. As a consequence, substellar companions that are cooler, less massive and at smaller separations can now be found than what was possible a few years ago. Examples of such detections are 2M 1207 B (Chauvin et al. 2005), and SCR 1845 B (Biller et al. 2006). Both of these objects in fact have planetary mass solutions within their error bars, but it should be stressed that these error bars are based on theoretical mass-luminosity relationships that are, so far, poorly calibrated. A further discussion regarding such theoretical models is performed in Sect. \ref{sec_limits}.

For a definitive detection of an extrasolar planet through direct imaging, one should preferably image an object which both has a low enough mass to be classified as such, and being close enough to its star that its actual mass can be determined by dynamical methods within a reasonable time frame. A particularly promising candidate system in this regard is $\epsilon$ Eri. A candidate planetary companion to the star $\epsilon$ Eri has been detected by radial velocity measurements (Hatzes et al. 2000). While the radial velocity signature by itself could in principle also be interpreted as being a result of the strong magnetic activity of $\epsilon$ Eri, it is important to note that if this was the case, one should also expect variations in the Ca II H\&K emission with the same periodicity as the radial velocity signal (see Baliunas et al. 1995). Since no such correlation could be found, Hatzes et al. (2000) concluded that a planetary companion was the most probable cause of the observed radial velocity variations. 

Subsequently, astrometry presented by Benedict et al. (2006) yielded further evidence for a planetary companion. By combining HST FGS astrometry with MAP astrometry (Gatewood 1987), and the radial velocity data from Hatzes at el. (2000) along with additional radial velocity measurements, Benedict et al. (2006) found consistent and statistically significant evidence for a planetary companion. Unfortunately, since the HST FGS astrometry has ceased operation, the astrometry does not cover the full orbit, which would further strengthen the conclusion of the existance of $\epsilon$ Eri b. Still, with the two different lines of evidence pointing to the presence of a planetary companion, $\epsilon$ Eri b is a significantly stronger candidate than the majority of extrasolar planet candidates known to date. On this note, it should also be pointed out that based on HIPPARCOS data, Wielen et al. (1999) mark $\epsilon$ Eri as a $\Delta \mu$ binary at the limit of detectability, which is consistent with a planetary mass companion around the expected separation of $\epsilon$ Eri b, yielding yet another piece of independent evidence for a planetary companion. Hence, throughout this paper we will assume that the planet exists with the orbital configuration given in Benedict et al. (2006). In the event that $\epsilon$ Eri b, despite the evidence indicating otherwise, should not exist, the detection limits for other substellar companions around $\epsilon$ Eri as a function of separation from the primary can be read out from Figs. \ref{ts_h} and \ref{ts_h_vlm}.

Benedict et al. 2006 give a mass of 1.55 $M_{\rm jup}$ for $\epsilon$ Eri b. The system is located just 3.2 pc away, making it the nearest extrasolar planetary system known to date. In addition, it has been estimated that the system is relatively young (at least within 0.1 to 1 Gyr, see Sect. \ref{sec_limits}), which is preferable from an observational point of view since at younger ages, the brightness contrast between the primary and secondary is smaller. Despite its youth, for the predicted age of this system and the measured mass of the planet, $\epsilon$ Eri b is expected to be significantly cooler than 700 K (based on the models of Baraffe et al. 2003), implying that it will exhibit strong methane absorption, which can be taken advantage of through spectral differencing. For these reasons, we have performed a multi-epoch observing campaign with state-of-the-art equipment and methods in an attempt to directly image $\epsilon$ Eri b. Very sensitive searches for planetary mass companions to $\epsilon$ Eri have been performed previously with Keck (Macintosh et al. 2003) and Spitzer (Marengo et al. 2006), but these searches aimed at the detection of more distant companions, and were sensitive only to separations of several arcseconds. Since the projected separation of $\epsilon$ Eri b as suggested by the dynamical measurements is always smaller than about 1.7'', the survey presented here is the first one with a hypothetical possibility to detect this companion. 

In this paper, we present the results of these observations in combination with astrometric data, and discuss the limits this implies for the properties of $\epsilon$ Eri b. We also analyze the NACO-SDI data with the aim to find appropriate strategies for how to efficiently defeat the residual noise, which is a complex mixture of correlated and uncorrelated, dynamic and quasi-static noise contributions.

\section{Observations}

\label{sec_obs}

Our imaging observations of $\epsilon$ Eri were taken at four different epochs: (1) In August of 2003 during a commissioning run, (2) in September of 2004, (3) in August of 2005 and (4) in December of 2005 and January of 2006. All the images were taken with the NACO adaptive optics system at VLT (UT4) on Paranal, Chile. 

At all epochs we used a combination of two differential imaging techniques: simultaneous spectral differential imaging (SDI) and angular differential imaging (ADI). For the purpose of SDI, narrow-band images were taken simultaneously in filters that we will refer to as F1, F2, and F3 which correspond to wavelengths of 1.575 $\mu$m, 1.600 $\mu$m, and 1.625 $\mu$m, respectively. For ADI, these observations were repeated at two different rotator angles (0$^{\circ}$ and 33$^{\circ}$, respectively). The rationale behind these techniques is briefly discussed in Sect. \ref{sec_datared}. The epoch 4 data are spread over 3 weeks, which is a short period compared to the duration of the orbit -- however, judging from the current best-fit astrometry, it appears that the companion is close enough to periastron at epoch 4 that it should be expected to move about 34 mas during this period. Hence, we should expect a slightly elongated image of the planet in the epoch 4 data due to its orbital motion in the 3 week period over which the data have been obtained.

For estimating the Strehl ratios of each of the frames, we used the coherent energy, which is a quantity that is measured automatically during all observing runs and stored in the image header. The details of conversion between coherent energy and Strehl are given in Fusco et al. (2004). For Strehl ratios higher than about 1 \% in H-band (10 \% in K-band), the coherent energy is a good approximation (the standard deviation is about 7.2 \%) to the Strehl ratio at a wavelength of $\lambda _1 = 2.166$ $\mu$m (i.e., the measured coherent energy corresponds to the Strehl ratio at a wavelength of 2.166 $\mu$m, regardless of at which wavelength the measurements are taken). We can rescale this quantity to our working wavelength $\lambda _2$ by using the definition of coherent energy and the Mar\'echal equation, giving:

\begin{equation}
S_2^{\star} = \exp(\ln S_1^{\star} (\lambda _1 / \lambda _2)^2)
\end{equation}

where $S_1^{\star} $ and $S_2^{\star}$ are the coherent energies at wavelengths $\lambda _1$ and $\lambda _2$, respectively (note that coherent energies and Strehl ratios are usually given in percentages, but in this equation they must be put in as fractions of one, i.e., if the coherent energy is 30 \%, the number 0.3 should be used). The average Strehl and other observation parameters are shown in Table \ref{tab1} for all four epochs.

\begin{figure}[htb]
   \centering
   \includegraphics[width=8.0cm]{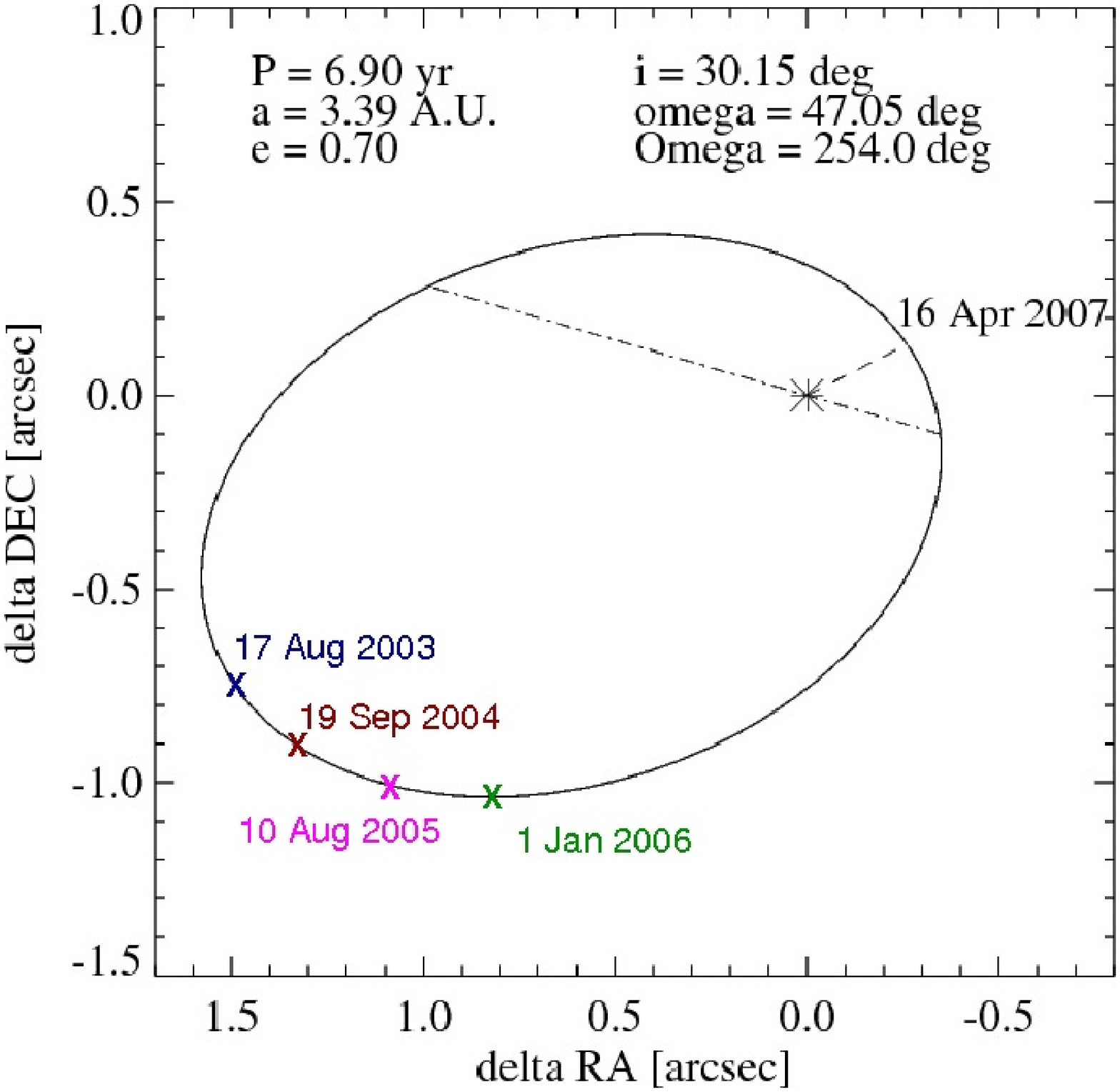}
\caption{Best-fit astrometrical orbit of $\epsilon$ Eri b around its parent star. The solid line marks the orbit, the dashed line shows the periastron, and the dash-dotted line shows the nodes. The approximate poisitions of the planet at each epoch of observation are also shown.}
\label{cart_orb}
    \end{figure}

To enhance the detectability and strengthen the reliability of any companion that might be found in the images, we have incorporated astrometric data from mainly HST, complemented by MAP ground-based measurements. The astrometry is discussed in Benedict et al. (2006). The best-fit orbit is shown in Fig. \ref{cart_orb}.

The specifics of the NACO system are detailed in Lenzen et al. (2003) and Rousset et al. (2003). In short, NACO is located in one of the Nasmyth foci of the VLT. It rotates around one axis to compensate for the internal flexing of the instrument due to the changing gravity vector with respect to the focal plane of the camera (CONICA). The calibration of static aberrations is described in Blanc et al. (2003) and Hartung et al. (2003). No coronograph or other attenuation device was used for these observations. Flexures within the adaptive optics path of the instrument are not perfectly compensated -- the rotation of the instrument causes small mis-alignments between the wavefront sensor subpupils and the deformable mirror. This leads to a Strehl ratio degradation, and most of the residual speckles, which are still present after SDI and ADI correction (see Fusco et al. 2005). Differential static aberrations, which are due to the different wavepaths at different wavelengths, are discussed in Brandner et al. (2004).

\section{Data reduction}

\label{sec_datared}

\begin{figure}[htb]
   \centering
   \includegraphics[width=8.0cm]{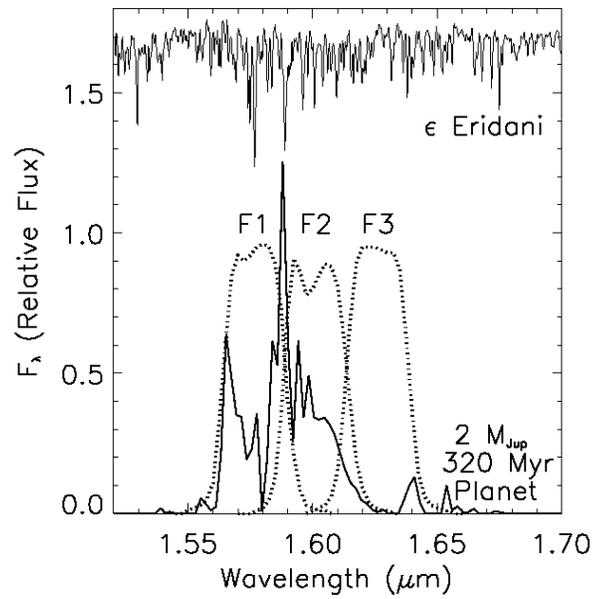}
\caption{Filter transmission curves of narrow-band filters F1, F2, and F3, along with the spectral distribution of $\epsilon$ Eri A from Meyer et al. (1998) and a theoretical spectrum from Burrows et al. (2003), similar to what would be expected from $\epsilon$ Eri b. The flux of $\epsilon$ Eri A is essentially uniform over the whole range, whereas the flux of the companion is strongly concentrated within the range of F1 and F2, according to theoretical models.}
\label{ee_filters}
    \end{figure}

We primarily used a dedicated SDI/ADI data reduction pipeline (see Kellner 2005, or Kellner et al., in preparation, for an extensive discussion -- also, a very similar reduction scheme is detailed in Biller et al. 2004) for reducing the data. Each frame was background subtracted, flat-fielded and filtered with a bad pixel mask. The SDI was performed by subtracting F3 from F1 and F2 separately, after rescaling to a common $\lambda$/D scale. The common rationale behind SDI is that cool enough objects (T-dwarfs and giant planets) exhibit methane absorption, which results in an absorption band starting at about 1.6 $\mu$m and stretching towards longer wavelengths. For a stellar object, on the other hand, the spectral continuum is rather constant over this spectral range. Thus, in a system of a star with a substellar companion, the companion will appear much brighter in a 1.575 $\mu$m (F1) image than in a 1.625 $\mu$m (F3) image, whereas the star will be equally bright in both frames. By subtracting the latter image from the former, the companion will therefore remain largely unaffected in the difference frame, whereas the primary will be mostly canceled out. If the narrow-band images are taken simultaneously as in this case, removal of the stellar PSF includes attenuation of the halo speckle noise, which otherwise is by far the dominant noise source in high-contrast, low-separation imaging. For an object as low-mass as Eps Eri b, an additional factor plays in, in that additional absorption will start to decrease the brightness in F1 for increasing ages, such that Eps Eri b becomes equally bright in F2 as in F1, and eventually even brighter in F2 according to the Burrows et al. (2003) model (see Fig. \ref{ee_filters}). Hence, we use both the F1-F3 and F2-F3 difference images, so as to be optimally sensitive over a wide range of ages. We refer to Racine et al. (1999) for a more detailed discussion on speckle noise and SDI.

All SDI frames for each angle were then co-added to make use of the full integration time. Finally, the ADI was performed by subtracting the 33$^{\circ}$ data from the 0$^{\circ}$ data. The idea behind the ADI technique is that the telescope and instrumentation give rise to static aberrations in the final image -- in particular, the SDI setup leads to non-common path aberrations since the light is split up for simultaneous imaging. However, when rotating the camera by e.g. 33$^{\circ}$ with all other set-ups being the same, a companion will rotate with respect to its primary by 33$^{\circ}$ in the resulting image, whereas the static aberrations should be unaffected. Thus, by subtracting two images at different angles, these aberrations will cancel out whereas the companion will remain with a very particular signature of one positive and one negative peak, at the same separation from the primary, but at a position angle differing by 33$^{\circ}$. The principle is also known as roll deconvolution, and has been frequently used for, e.g., the HST (see Mueller \& Weigelt 1987 and subsequent publications). ADI is also used by Marois et al. (2006), but with a somewhat different implementation, where images are taken at several different angles. Marois et al. (2006) get a noise reduction of about a factor of 5 for each image subtraction with such an implementation. Combining SDI and ADI with our implementation gives a noise reduction of 2 to 3 magnitudes, i.e. a factor of 6-16 improvement for each image subtraction (Kellner 2005). It would be an interesting experiment, as is suggested in Marois et al. (2006), to combine SDI with their implementation of ADI in order to possibly increase the sensitivity somewhat further.

While our three narrow-bands in principle allow for multi-wavelength image subtraction in the manner described by Marois et al. (2000), this cannot be applied in practice, due to the fact that static or quasi-static aberrations are present in the data, which were not considered in Marois et al. (2000). The static aberrations influence the $k$-factor derived in Marois et al. (2000) and prevent any increase of quality from this method. For future instrumentation with possibly smaller static aberrations, this technique may be highly interesting to add to the combination of differential imaging methods.

Given that the final output is strongly affected by the Strehl ratio of individual frames, it is not a given fact that co-adding as many frames as possible will necessarily add up to the best possible result. In some cases, it is instead preferable to de-select frames with a bad Strehl ratio if they do more harm than good to the final frame. We therefore performed a number of tests to determine the optimal selection of frames. This was done by sequentially (and cumulatively) de-selecting the frames with the lowest measured Strehl ratio and checking the quality. The quality criterion was a minimization of the average error in the area between 20 and 80 pixels away from the star, divided by the average Strehl ratio of the sample in order to take into account the fact that the brightness of a hypothetical companion PSF core would be proportional to the Strehl ratio. The error was quantified by the standard deviation in a 9x9 pixel area around each pixel. For epochs 1 and 2, we found that no substantial improvement could be gained by excluding frames from the full set. For the epoch 3 data, we found that a de-selection of the four worst frames (in terms of Strehl ratio) per angle gave the best overall quality, and therefore we used the resulting set for further analysis. For the epoch 4 data, a slight error during observation led to three more frames for the 33$^{\circ}$ data set than for 0$^{\circ}$. It is preferable that the number of frames is the same at both orientation, so that the noise impact is equal during subtraction. For this reason, we de-selected the three worst frames for 33$^{\circ}$ to begin with. Subsequent analysis in the same manner as for earlier epochs led to the conclusion that the best quality was reached by keeping all remaining frames in the final selection.

\section{Results and discussion}

\label{sec_results}

\subsection{Analyzing the images}

\label{sec_analyze}

\begin{figure*}[htb]
   \centering
   \includegraphics[width=16.0cm]{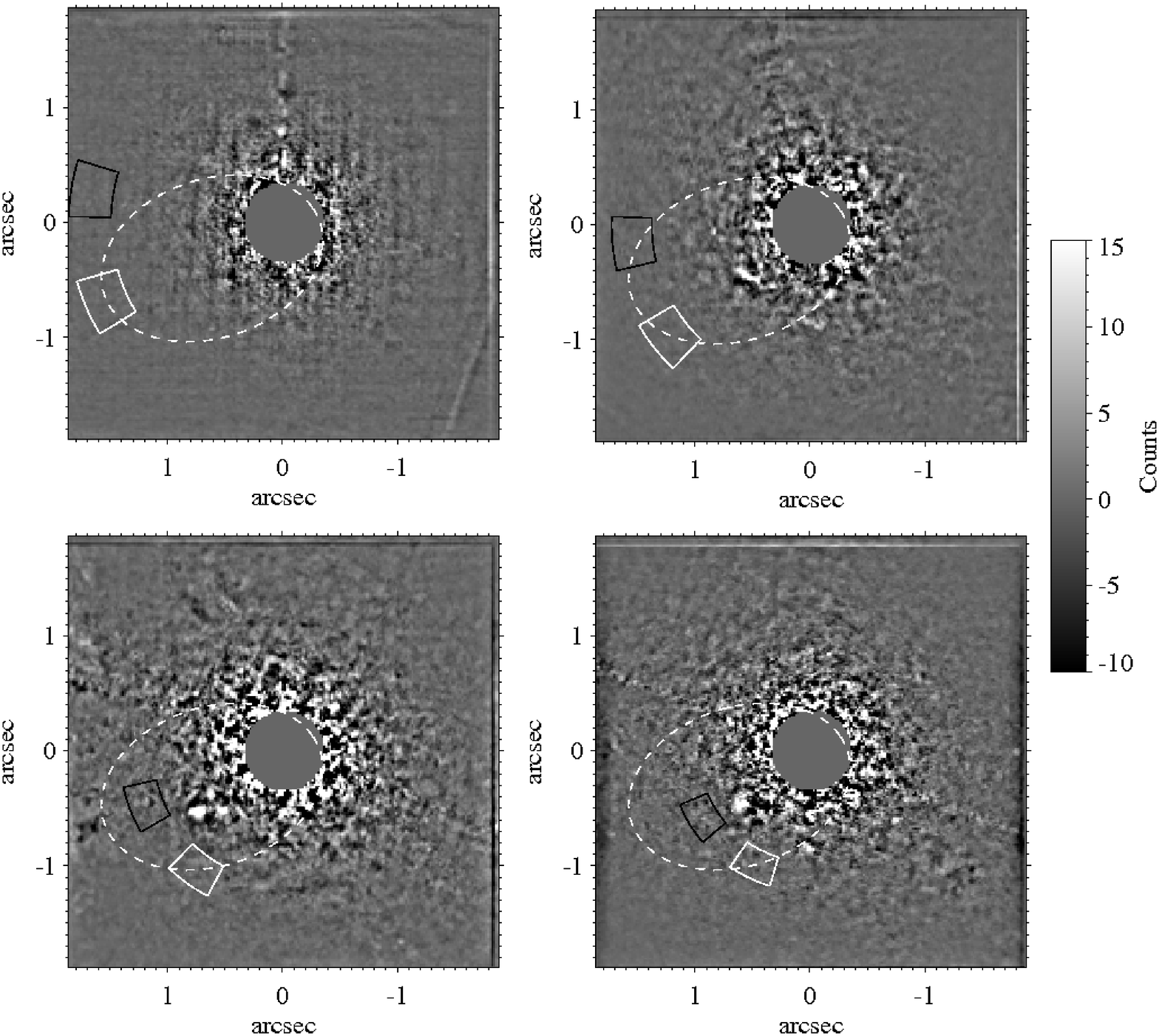}
\caption{The output F1-F3 images from each of the observations in sequence. Upper left: Epoch 1. Upper right: Epoch 2. Lower left: Epoch 3. Lower right: Epoch 4. The dotted line marks the best-fit orbit from astrometric and radial velocity data. The areas enclosed by white and black borders are error boxes for the expected positions of the bright and dark signatures of the companion, respectively. In all of the images, north is up, and east is to the left. All the counts are per pixel.}
\label{ee_epochs}
    \end{figure*}

\begin{figure*}[htb]
   \centering
   \includegraphics[width=16.0cm]{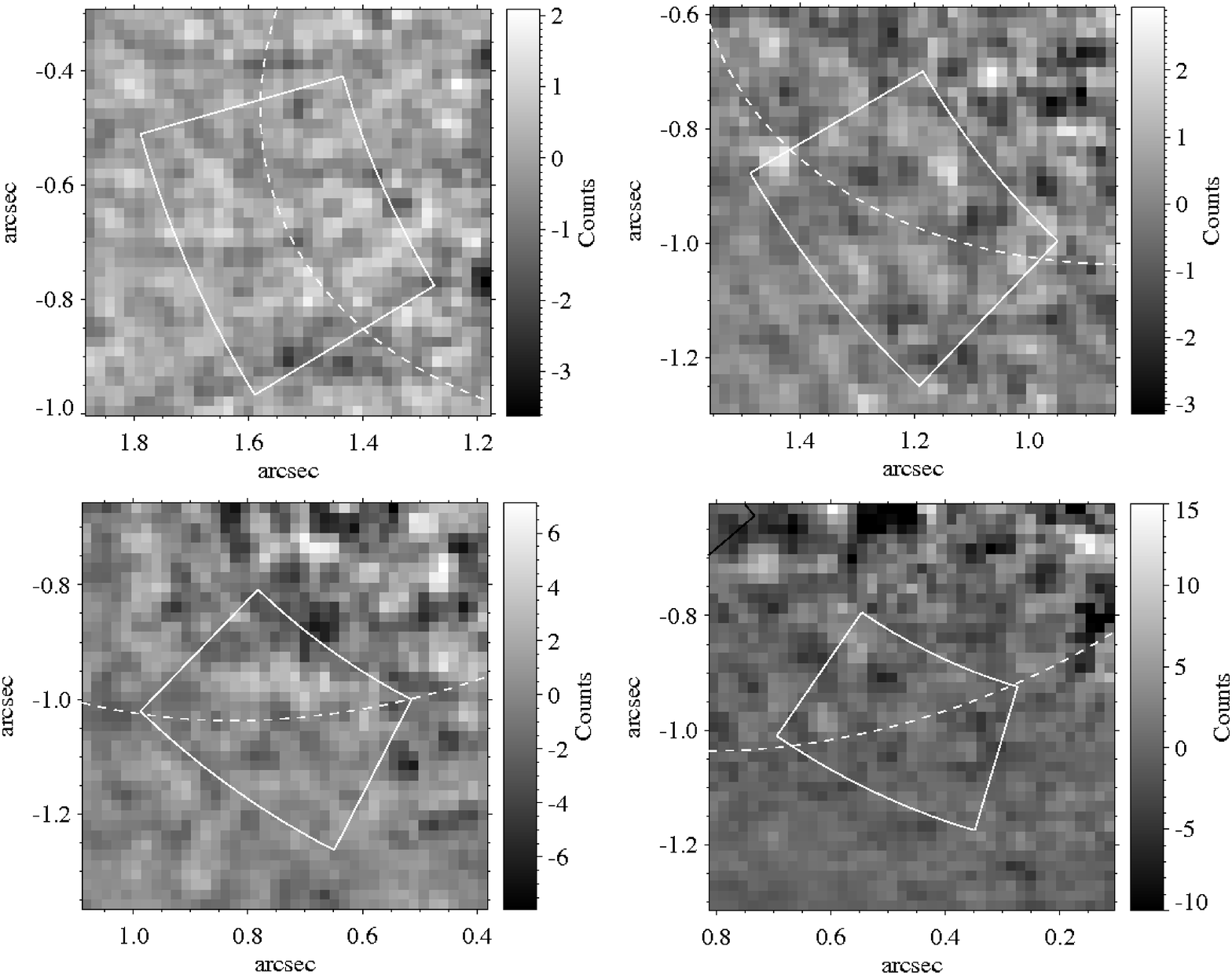}
\caption{Zoomed-in counterparts of the F1-F3 images from each epoch in Fig. \ref{ee_epochs}. The field of view is centered on the expected position of the companion.}
\label{ee_zooms}
    \end{figure*}

\begin{figure*}[htb]
   \centering
   \includegraphics[width=16.0cm]{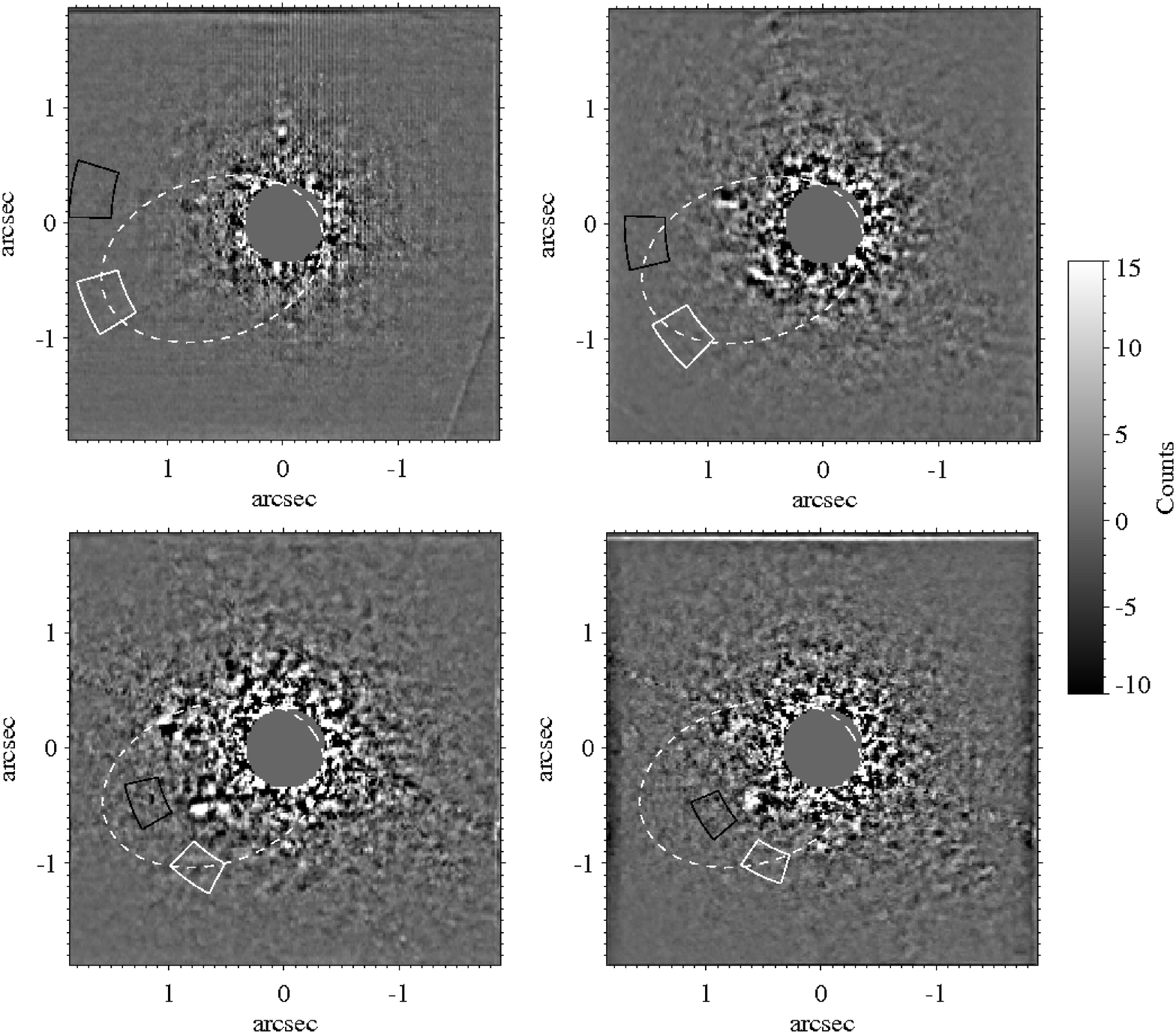}
\caption{The output F2-F3 images from each of the observations in sequence. Upper left: Epoch 1. Upper right: Epoch 2. Lower left: Epoch 3. Lower right: Epoch 4. The dotted line marks the best-fit orbit from astrometric and radial velocity data. The areas enclosed by white and black borders are error boxes for the expected positions of the bright and dark signatures of the companion, respectively. In all of the images, north is up, and east is to the left.}
\label{ee_vlms}
    \end{figure*}

\begin{figure*}[htb]
   \centering
   \includegraphics[width=16.0cm]{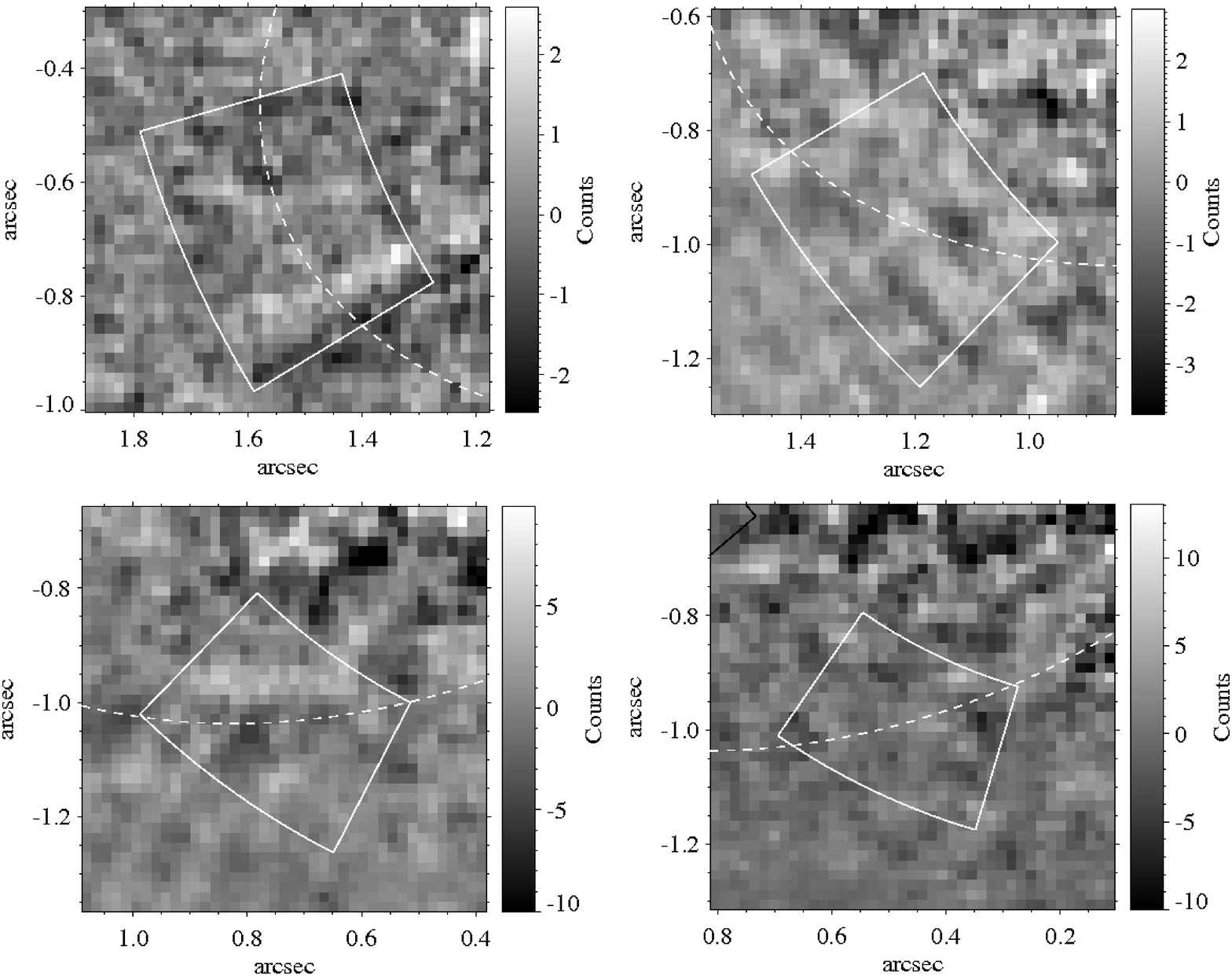}
\caption{Zoomed-in counterparts of the F2-F3 images from each epoch in Fig. \ref{ee_epochs}. The field of view is centered on the expected position of the companion.}
\label{ee_zoom_vlms}
    \end{figure*}

The reduced images are shown in Fig. \ref{ee_epochs} for F1-F3, and Fig. \ref{ee_vlms} for F2-F3. To enhance the conditions for visual inspection in the interesting areas, the central areas have been set to zero. While there is meaningful information in most of these areas, the fluctuations are much larger, and detection of a planetary mass companions is therefore not possible there. Also, a few of the central pixels are normally saturated in the raw images, hence no meaningful information is available at the very center (within about 5 pixels). Zoomed-in versions of the most interesting areas from an astrometric point of view are shown in Figs. \ref{ee_zooms} and \ref{ee_zoom_vlms}. The main problem of finding a faint companion in the final data is readily seen in those images: the correlated residual speckle noise forms a vast number of coherent structures in the image space, which mimic the appearance of a physical companion. The ADI is however a great help in this regard. A real companion has to leave an imprint of one positive and one negative structure in the image, where both structures are at the same separation from the center of the stellar PSF (the position of which must of course be saved during the data reduction, since the PSF is canceled out in the difference images). The negative peak has to be separated from the positive one by 33$^{\circ}$ clockwise. Still, the centers of the respective companion peaks, as well as the center of the stellar peak, can not be determined with infinite precision. Thus, there will remain several false positives in the images since the speckles are common enough that a negative speckle will, by chance, end up close enough to the right position relative to a positive speckle in several cases. 'Negative' and 'positive' speckle in this context denotes a coherent residual structure, from atmospheric or instrumental aberrations, which is brighter in one frame than the other during either of the differencing stages -- i.e., in the difference image $a-b$, a speckle becomes positive if it is brighter in $a$ than in $b$, and vice versa.

\begin{figure}[htb]
   \centering
   \includegraphics[width=12.0cm]{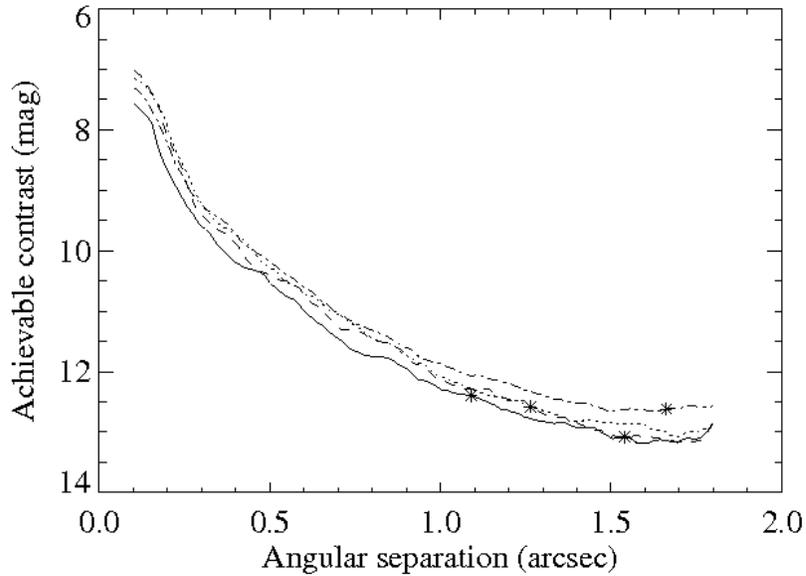}
\caption{$3 \sigma$ detection limits for our images of $\epsilon$ Eri as a function of radial angular separation, based on local standard deviations. The limits are based on the narrow-band F1 brightness contrast (for H-band contrasts, see Figs \ref{ts_h} and \ref{ts_h_vlm}). The dash-dotted line is the epoch 1 data, the dotted line is the epoch 2 data, the dashed line is the epoch 3 data, and the solid line is the epoch 4 data. The star on each curve represents the expected angular separation based on the dynamical measurements. It can be seen that even though the epoch 4 data has the highest overall sensitivity, the smaller expected separation of $\epsilon$ Eri b leads to a worse detection limit than for the other epochs. The range within 0.1 arcsec, where saturation occurs in some frames, has been set to zero.}
\label{three_sigma}
    \end{figure}

\begin{figure}[htb]
   \centering
   \includegraphics[width=12.0cm]{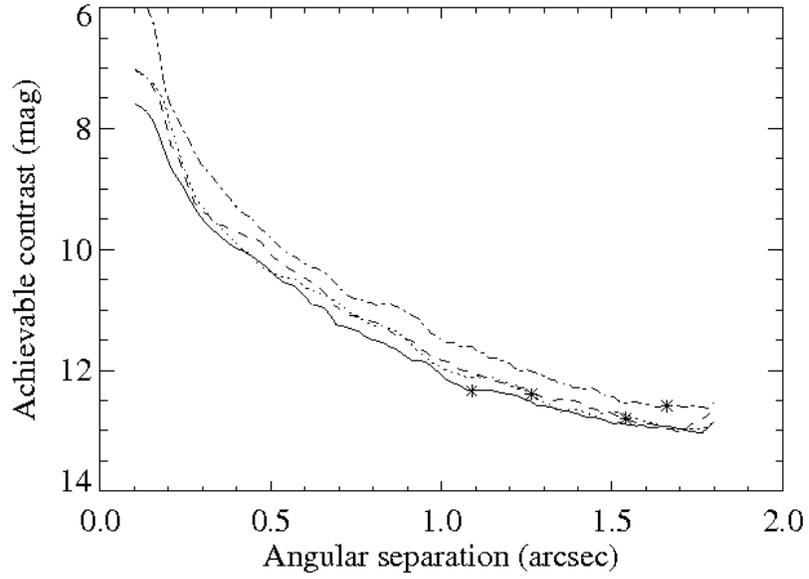}
\caption{Same as Fig. \ref{three_sigma}, but for F2-F3 instead of F1-F3. The sensitivity is somewhat worse (by a few tenths of a magnitude) in this case, probably due to a worse quality of the F2 sub-frame.}
\label{ts_vlm}
    \end{figure}

\begin{figure*}[htb]
   \centering
   \includegraphics[width=12.0cm]{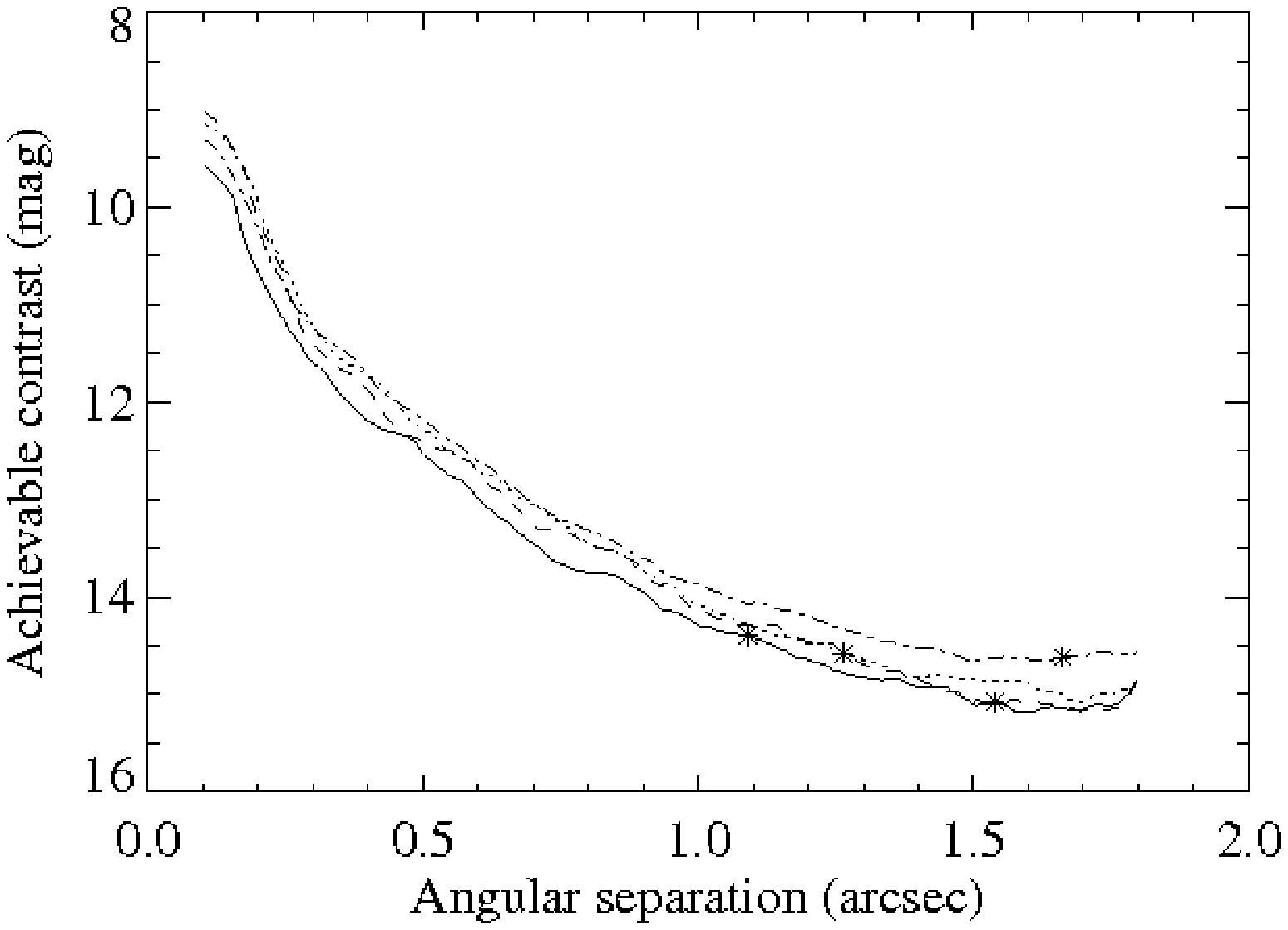}
\caption{H-band brightness contrasts for a 3$\sigma$ detection, based on the calculated offsets and the F1 detection limits (see Fig. \ref{three_sigma}), assuming an age of 100 Myr. The stars mark the expected position, based on the current best-fit astrometry, of the companion at each epoch.}
\label{ts_h}
    \end{figure*}

\begin{figure*}[htb]
   \centering
   \includegraphics[width=12.0cm]{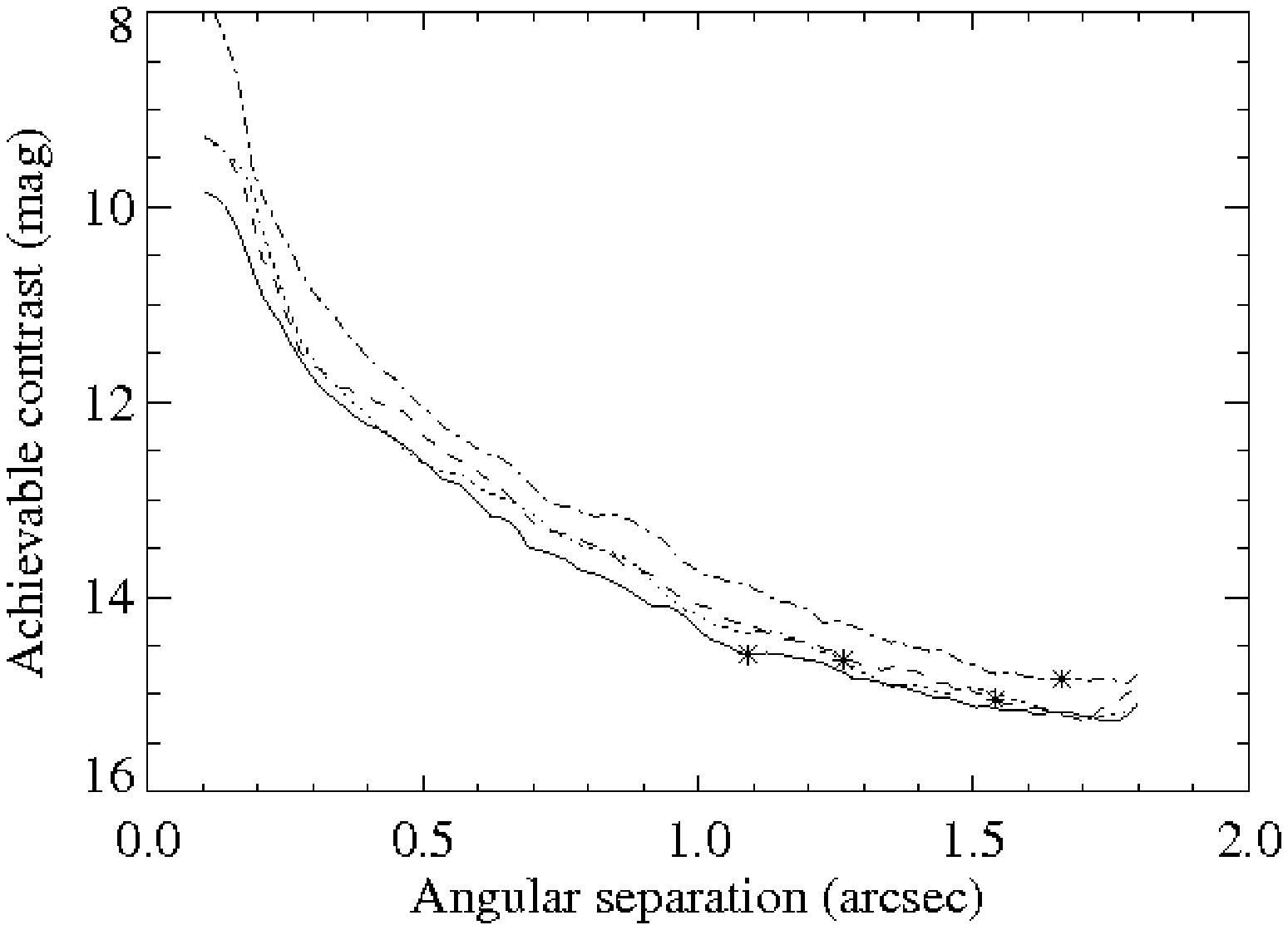}
\caption{H-band brightness contrasts for a 3$\sigma$ detection, based on the calculated offsets and the F2 detection limits (see Fig. \ref{three_sigma}), assuming an age of 1 Gyr. The stars mark the expected position, based on the current best-fit astrometry, of the companion at each epoch.}
\label{ts_h_vlm}
    \end{figure*}

The $3 \sigma$ narrow-band detection limits per pixel of each epoch are shown in Fig. \ref{three_sigma} for F1-F3, and Fig. \ref{ts_vlm} for F2-F3. The limits are based on the median of the statistical errors at the various radii. We will discuss what they correspond to physically in Sect. \ref{sec_limits}. It can be seen by comparing those figures, as well as the visual quality of the images, that F1-F3 produces somewhat higher qualities than F2-F3 on average. With respect to the images, the limits are such that several candidates exist with fluxes above the $3 \sigma$ limit. However, as we have already alluded to, this is not sufficient to claim a detection. The detection limits are useful as they give a general view of the sensitivity of the data, but when dealing with speckle noise, it is necessary to have additional constraints to the $3 \sigma$ threshold that can be used for detection when limited by uncorrelated noise. This is due to the fact that the residual errors are not Gaussian, and hence $3 \sigma$ does not correspond to the well known 99.7 \% detection confidence. One such constraint can be to demand that the candidate clearly dominates the speckle noise, i.e., to set an extremely high threshold such that no single speckle could be bright enough to mimic the appearance of such a companion. Another way to constrain the data is to incorporate a priori information about the properties of the companion that a given candidate has to match (as has, e.g., been done for GQ Lup b, see Janson et al. 2006). Since there is no candidate that dominates the flux by an extreme amount (though some candidates are of course stronger than others in this regard, see e.g. Kellner et al., in preparation), we try the latter alternative.

As we have mentioned, radial velocity and astrometry data exist that we can use to determine the orbit of $\epsilon$ Eri b, and thus its position relative to $\epsilon$ Eri at any given epoch. Using the orbital parameters, and using a mass estimate of Benedict et al. (2006) for $\epsilon$ Eri ($0.83 M_{\rm sun}$), we find separations and position angles for $\epsilon$ Eri b as compiled in Table \ref{tab3} for each epoch at which our images were taken. The results are also overplotted in Figs. \ref{ee_epochs} to \ref{ee_zoom_vlms}. Errors in separation and position angles are derived by generating $10^4$ orbits with random errors for the orbital parameters set by the values given for the errors in Benedict et al. (2006), and calculating the resulting standard deviations in separation and position angle at each of the expected positions of $\epsilon$ Eri b. The error boxes can be used for excluding a large amount of false positives. Since we have multiple epoch data, we can also in principle acquire a robust detection of a real object -- if a candidate shows up with its positive and negative signatures in the right places during all epochs, we can calculate the probability that this would happen by chance with speckles, which will give us a meaningful statistical basis on which to confirm (or not confirm) the detection of a companion. The statistical analysis could for instance be done in the following way: Within a circular zone with inner and outer radii set by the known separation from astrometry with error bars, a number count is done of positive and negative speckles above a certain threshold. Based on this and the area of the zone, we can calculate the probability that a positive speckle ends up within its astrometric error bars, and that a negative speckle simultaneously ends up within its corresponding error bars (which is a sub-zone within the circular area, limited by the error bars in position angle). This probability can be calculated for each epoch, and by multiplying these probabilities, a final probability is acquired which can be required to be, e.g., less than 1 \%. In our case, the error bars are not sufficiently well-constrained that a meaningful analysis can be done in such a manner (i.e., the areas are large enough that speckles can not be excluded with sufficient confidence). This can however be significantly improved upon with further astrometric monitoring.

In summary, we do not detect any sufficiently significant candidates in the data to claim a detection of $\epsilon$ Eri b, though with additional dynamical data, the images may still be useful in this regard. 

\subsection{Detection limits}

\label{sec_limits}

In the previous section, we presented the statistical errors of each epoch (Figs. \ref{three_sigma} and \ref{ts_vlm}). We can use these errors to estimate detection limits in observational terms (narrow-band brightness contrast). Then, by inferring theoretical evolutionary models, we can formulate them in more physically relevant terms. The observational narrow-band detection limits are, of course, directly available from the figures. It can be seen that overall, epoch 4 in F1-F3 provides the most sensitive data. In this case, at 0.5'', a contrast of about 10.5 mag can be reached between primary and secondary, and at 1.0'', a contrast of almost 12.5 mag can be reached. At about 1.5'' and outwards, we reach a contrast of 13 mag. However, if we take the best-fit astrometry into account, we see that for $\epsilon$ Eri b, epoch 4 actually provides the least sensitive data point, with a contrast of about 12.4 mag. Epochs 1 and 3 are somewhat better with about 12.6 mag in both cases. The most sensitive measurement according to the astrometry is clearly the epoch 2 data, providing a contrast of $\sim 13.1$ mag. 

The astrometry provides a unique mass for $\epsilon$ Eri b of about $1.55 M_{\rm jup}$ (Benedict et al. 2006). To get a handle on whether we could expect to detect the planet in our data, we can translate the mass into a brightness using the theoretical mass-luminosity relationships of Baraffe et al. (2003) as a function of age. In this context, we wish to carefully remind the reader that such relationships are hugely uncertain for such low-mass objects, in particular for young ages. Indeed, comparison of the measured brightness and the dynamical mass of the young star-BD boundary object AB Dor C seems to imply that the theoretical models overestimate the luminosity corresponding to a given mass for such objects (see Close et al. 2005). However, this is based on an age estimate which has been questioned, and is a matter of discussion (see e.g. Luhman et al. 2005 and Janson et al. 2006). Also, there may be differences between properties of objects undergoing significant accretion, and objects which do not. Accretion is not considered in the Baraffe et al. (2003) models, but is a fundamental mechanism in the case of planet formation by core accretion. Marley et al. (2006) present models which do take this effect into account, and find that they differ drastically in predicted properties from collapse without accretion. However, for objects near $1 M_{\rm jup}$, such as $\epsilon$ Eri b, the discrepancy has vanished already at $\sim 10$ Myr. In any case, we assume that the Baraffe et al. (2003) model applies, which gives results summarized in Table \ref{tab2}.

\begin{figure*}[htb]
   \centering
   \includegraphics[width=8.0cm]{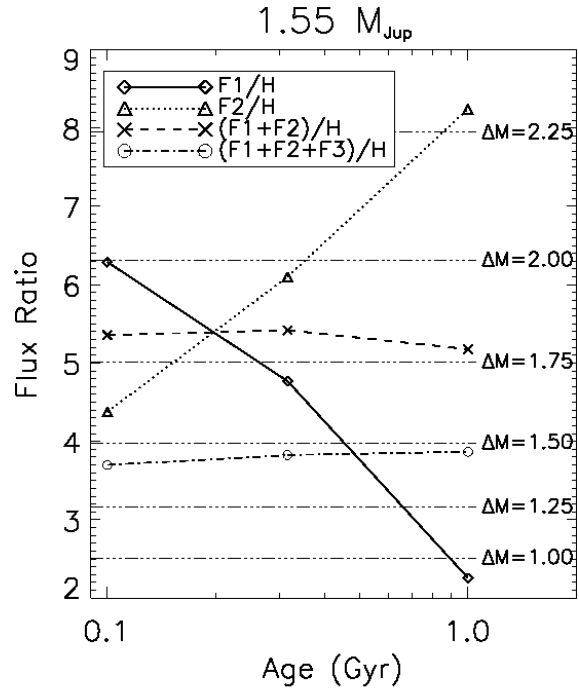}
\caption{Calculated narrow-band to H-band offsets for $\epsilon$ Eri b, based on the Burrows et al. (2003) model. Also plotted are the offsets that can be expected if the narrow-band images are averaged instead of differenced.}
\label{ee_burrows}
    \end{figure*}

For a fair comparison with our achieved contrasts, we translate the narrow-band contrasts given above into H-band contrasts. Such a procedure was first presented for T-dwarfs by Biller et al. (2006). However, since we know the mass of $\epsilon$ Eri b, we can do a much more specific analysis for this case. Using spectral models of Burrows et al. (2003), we calculate the offset $\Delta _{\rm mag}$ between F1 and H as:

\begin{equation}
\Delta _{\rm mag} = -2.5 \log _{10} (Q)
\end{equation}

\begin{equation}
Q = {{\int f_{\lambda} g_{\rm F1} d \lambda \over \int F_{\lambda} g_{\rm F1} d \lambda} \over {\int f_{\lambda} g_{\rm H} d \lambda \over \int F_{\lambda} g_{\rm H} d \lambda}}
\end{equation}

where $f_{\lambda}$ is the spectrum of the planet, $F_{\lambda}$ is the spectrum of the star, $g_{\rm F1}$ is the filter transmission of F1, and $g_{\rm H}$ is the filter transmission of H. An equivalent equation is valid for F2. The resulting offsets are plotted in Fig. \ref{ee_burrows} for ages of 100 Myr to 1 Gyr. It is clear that for young ages, F1-F3 is better suited for finding an object such as $\epsilon$ Eri b, whereas for older ages, F2-F3 is more appropriate. We show the calculated H-band contrasts in F1-F3 at 100 Myr in Fig. \ref{ts_h}, and in F2-F3 at 1 Gyr in Fig. \ref{ts_h_vlm}. The contrasts reached at the expected separation of Eps Eri range from about 14.5 mag (epoch 4) to about 15.1 mag (epoch 2), and imply that we could expect to detect $\epsilon$ Eri b with $3 \sigma$ confidence if the age is close to 50 Myr or younger, if the models are to be trusted. With an H-band brightness of 1.9 mag for the primary, and a distance modulus of about 2.5 mag, the epoch 2 data leads to a limiting absolute brightness of 19.5 mag for $\epsilon$ Eri b, though for ages between 100 Myr and 1 Gyr, the narrow-band to H-band offset is smaller, such that the minimum brightness limit is 19.1 mag for some ages in that regard.

Age estimates of $\epsilon$ Eri in the literature are quite divergent (see e.g. Song et al. 2000, Fuhrmann 2003, Decin et al. 2003, Saffe et al. 2005, and Di Folco et al. 2004), but seem to consistently yield ages larger than 100 Myr, and smaller than 1 Gyr. Thus, we conclude that we should not expect to detect the planet by $3 \sigma$ in any of the images. We note, however, that there may be other aspects to the problem that are not included in the above reasoning. Aside from that the models may mis-predict the brightness by an unknown factor due to the uncertain initial conditions, the brightness could also be affected by factors that are not included in the evolutionary models of Baraffe et al. (2003). A potentially interesting factor in this regard is interaction between the planet and a remnant debris disk. A debris disk has indeed been observed around $\epsilon$ Eri (see e.g. Greaves et al. 1998). Frequent collisions between the planet and the planetesimals in the disk would heat the outer atmosphere of the planet, temporarily leading to a substantial brightening. Since the magnitude of the effect depends on the frequency of collisions, and the conditions of the disk are poorly known, the magnitude of this effect is however difficult to determine.

Finally, we note that when surveying for the planets with constrained astrometry over several epochs as described in Sect. \ref{sec_analyze}, the $3 \sigma$ condition of the companion flux with respect to the source becomes less meaningful. Since in that case, the probability threshold is set by number counts of speckles, a source can be detected with a sufficient confidence in total, even though it may be less bright than the threshold set for a $3 \sigma$ detection for brightness within a single frame. As an example, we hypothesize that in each of four images, a 2$\sigma$ signature shows up within well-constrained astrometric error bars with both a signature of positive and negative counts in the right places. In none of the single cases, a detection can be claimed with any significant probability. However, we now assume that the probability of a $\ge 2 \sigma$ speckle ending up in the right places of a single image by chance is the same for all images, and can be estimated to be, say, 10 \%. The events in the four different images are independent, hence the total probability that the detection is false is $0.1^4 = 10^{-4}$. Hence in such a situation, a detection could be claimed with a sufficient confidence in total. 

\subsection{Error analysis}

\label{sec_error}

The noise in the final images is a complex mixture of dynamic and quasi-static, correlated and uncorrelated noise, with different relative impacts in different parts of the image. Correlated noise has a greater relative impact at small angular separations from the star, whereas the opposite is true for uncorrelated noise. The relative importance of these noise sources vary with varying observing conditions, such as the seeing. An analytical description of dynamic speckle noise versus uncorrelated noise sources (photon, read, and sky noise) is given in Racine et al. (1999), but quasi-static speckle noise, which is a major contributor to the noise in real applications, is not considered there. Other approaches for dealing with speckle noise based on, e.g., Goodman (1975) exist -- for instance, in Aime \& Soummer (2004). A method for using such an approach in practice is detailed in Fitzgerald \& Graham (2006). However, this methodology can not be well applied to NACO-SDI for atmospherical speckle noise, since it relies on getting a very large amount of very short exposures for statistical analysis. In the case of NACO-SDI it is essential to keep the integration time per exposure as large as possible, in order to minimize overhead time and read noise. For quasi-static speckle noise with timescales of a few seconds or larger, a similar technique could in principle be applied in the future, if the observing strategy is adapted appropriately. In general, the complexity of the noise makes it difficult to estimate a priori the observing conditions needed to reach a certain sensitivity for a certain source, when planning surveys for, e.g., extrasolar planets. 

In three of our epochs (2, 3, 4), we observe the same source, with the same instrument, the same detector and very similar observing strategies. Thus, we have a rather large amount of data where we can empirically test the quality of our data as a function of observing conditions, where all other parameters can be kept rather constant. Here we will perform an analysis of the normalized error $e$ as a function of Strehl ratio, and integration time: $e = \xi / S$ where $\xi$ is the average error, and $S$ is the Strehl ratio. The reason we divide by the Strehl ratio is the same as in Sect. \ref{sec_datared}: the Strehl ratio is proportional to the signal strength of a companion, and hence $e \sim SNR^{-1}$. Since the noise properties will vary with radial distance from the center of the remnant stellar PSF, we examine $e$ in four different zones separately: zone 1 is defined as the area between 0 and 19 pixels radially from the center, zone 2 as the area between 20 and 39 pixels, zone 3 between 40 and 59 pixels, and zone 4 between 60 and 79 pixels. Since the pixel scale of the NACO-SDI is 17.32 mas/pixel (see Brandner et al. 2004 for how this is determined), this corresponds to angular separations of about 0.0-0.3 arcsec for zone 1, 0.3-0.7 arcsec for zone 2, 0.7-1.0 arcsec for zone 3, and 1.0-1.3 arcsec for zone 4.

\subsubsection{Strehl ratio dependence}

\label{sec_strehl}

\begin{figure*}[htb]
   \centering
   \includegraphics[width=16.0cm]{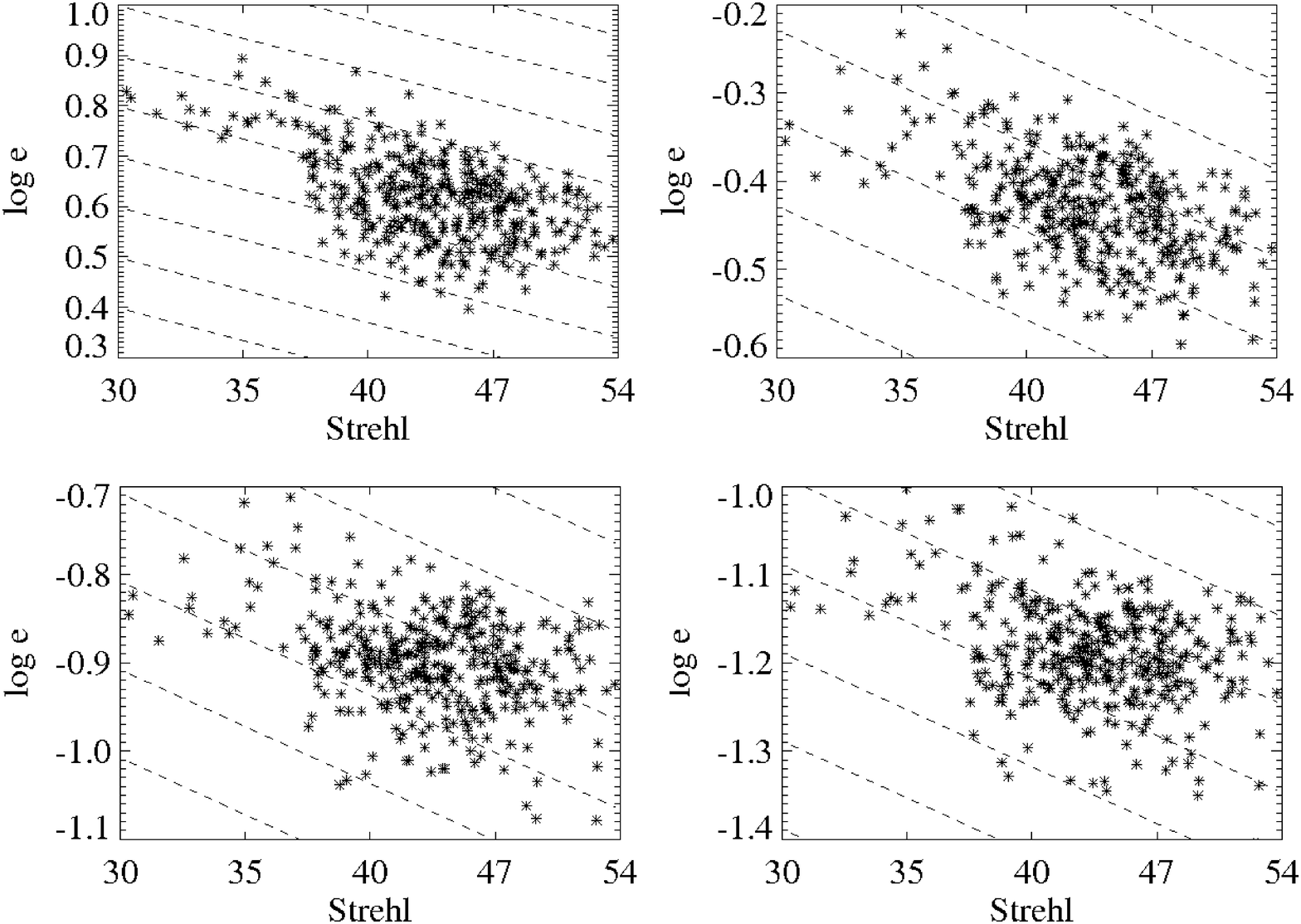}
\caption{The logarithmic Strehl-normalized error $e$ as a function of Strehl ratio for the epoch 2 data (stars). Upper left: Zone 1 (0-19 pixels away from the center). Upper right: Zone 2 (20-39 pixels). Lower left: Zone 3 (40-59 pixels). Lower right: Zone 4 (60-79 pixels). The dashed lines indicate a reference slope corresponding to $e \sim S^{-1}$. Since the range of Strehl ratios is small for this epoch, and the dispersion is rather large, the trend is not very easily seen in this case. However, note that the result is entirely consistent with epoch 3 (see Fig. \ref{ep3_strehls}). All the four zones appear to give very similar results. Note that since the dashed lines are equally spaced regardless of epoch and zone, it is easy to compare the dispersions. Note also that the x-axis is in logarithm scale.}
\label{ep2_strehls}
    \end{figure*}

\begin{figure*}[htb]
   \centering
   \includegraphics[width=16.0cm]{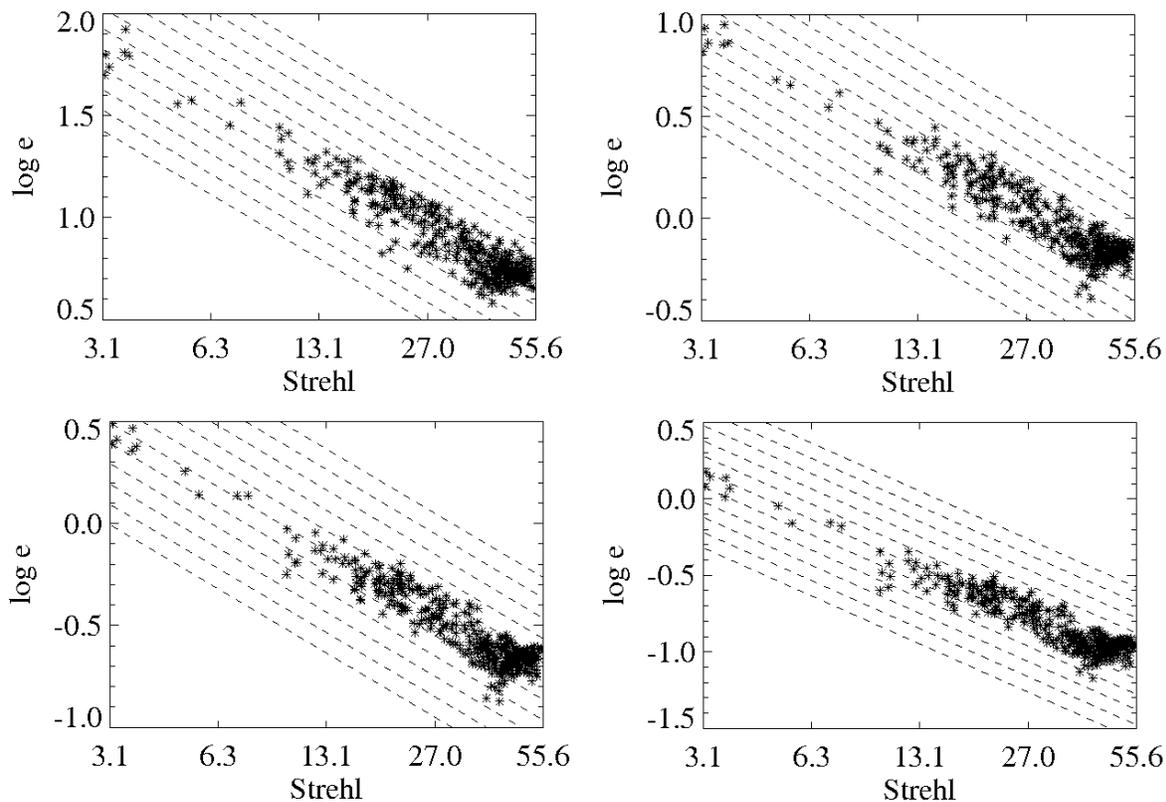}
\caption{The logarithmic Strehl-normalized error $e$ as a function of Strehl ratio for the epoch 3 data (stars). Upper left: Zone 1 (0-19 pixels away from the center). Upper right: Zone 2 (20-39 pixels). Lower left: Zone 3 (40-59 pixels). Lower right: Zone 4 (60-79 pixels). The dashed lines indicate a reference slope corresponding to $e \sim S^{-1}$. Despite the fact that the dispersion is equally large here as for the other epochs, the trend is particularly obvious for this case, since the Strehl ratios cover such a relatively wide range. All the four zones appear to give very similar results. Note that the x-axis is in logarithm scale.}
\label{ep3_strehls}
    \end{figure*}

\begin{figure*}[htb]
   \centering
   \includegraphics[width=16.0cm]{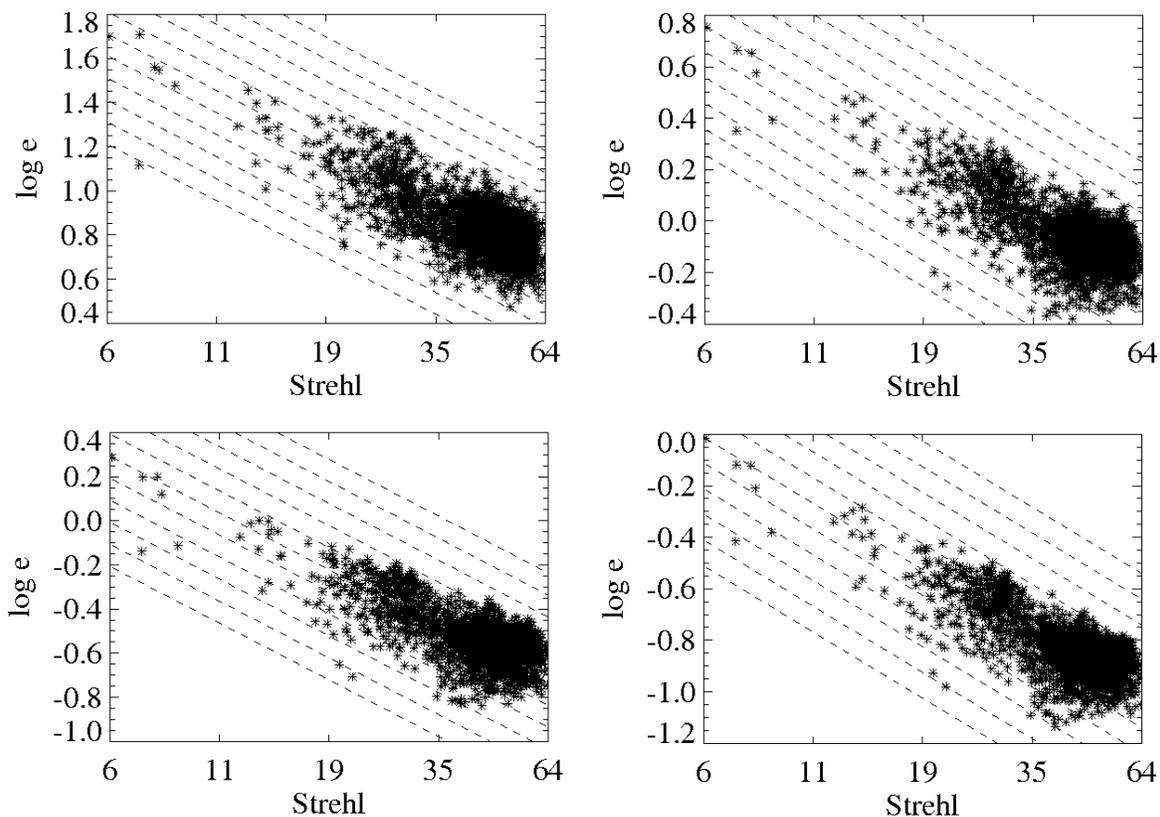}
\caption{The logarithmic Strehl-normalized error $e$ as a function of Strehl ratio for the epoch 4 data (stars). Upper left: Zone 1 (0-19 pixels away from the center). Upper right: Zone 2 (20-39 pixels). Lower left: Zone 3 (40-59 pixels). Lower right: Zone 4 (60-79 pixels). The dashed lines indicate a reference slope corresponding to $e \sim S^{-1}$. Thanks to the very large number of data points for this epoch, it is quite clear that the data follows the expected $e \sim S^{-1}$ trend. All the four zones appear to give very similar results. Note that the x-axis is in logarithm scale.}
\label{ep4_strehls}
    \end{figure*}

To examine $e$ as a function of Strehl ratio, every individual 0$^{\circ}$ frame within one epoch is coupled with every 33$^{\circ}$ frame within the same epoch, and the full data reduction is performed for each pair of frames. The average $e$ is then plotted against the average $S$ for each pair in each zone. The plots are shown in Figs \ref{ep2_strehls} through \ref{ep4_strehls}. The dispersion is large, but fortunately, we have a lot of data points, and so the trend is very clear: $e \sim S^{-1}$, i.e. $SNR \sim S$. 

As expected, the examination shows that Strehl ratio is an important parameter for optimizing the sensitivity when searching for substellar companions. The trend of $SNR \sim S$ is consistent for all three epochs, and over all four zones. Thus, doubling $S$ will generally lead to a doubling of the $SNR$, regardless of where in the image a hypothetical companion may be situated. Since our Strehl ratios are within the range of about 20 \% to 40 \% in H-band, it is of course not possible to predict whether this trend holds also for extremely high (or low) Strehl ratios. However, within the range of what can be reached with present instrumentation, it is clear that aiming for the highest possible Strehl ratio is indeed a good strategy.

\subsubsection{Integration time dependence}

\label{sec_time}

\begin{figure*}[htb]
   \centering
   \includegraphics[width=16.0cm]{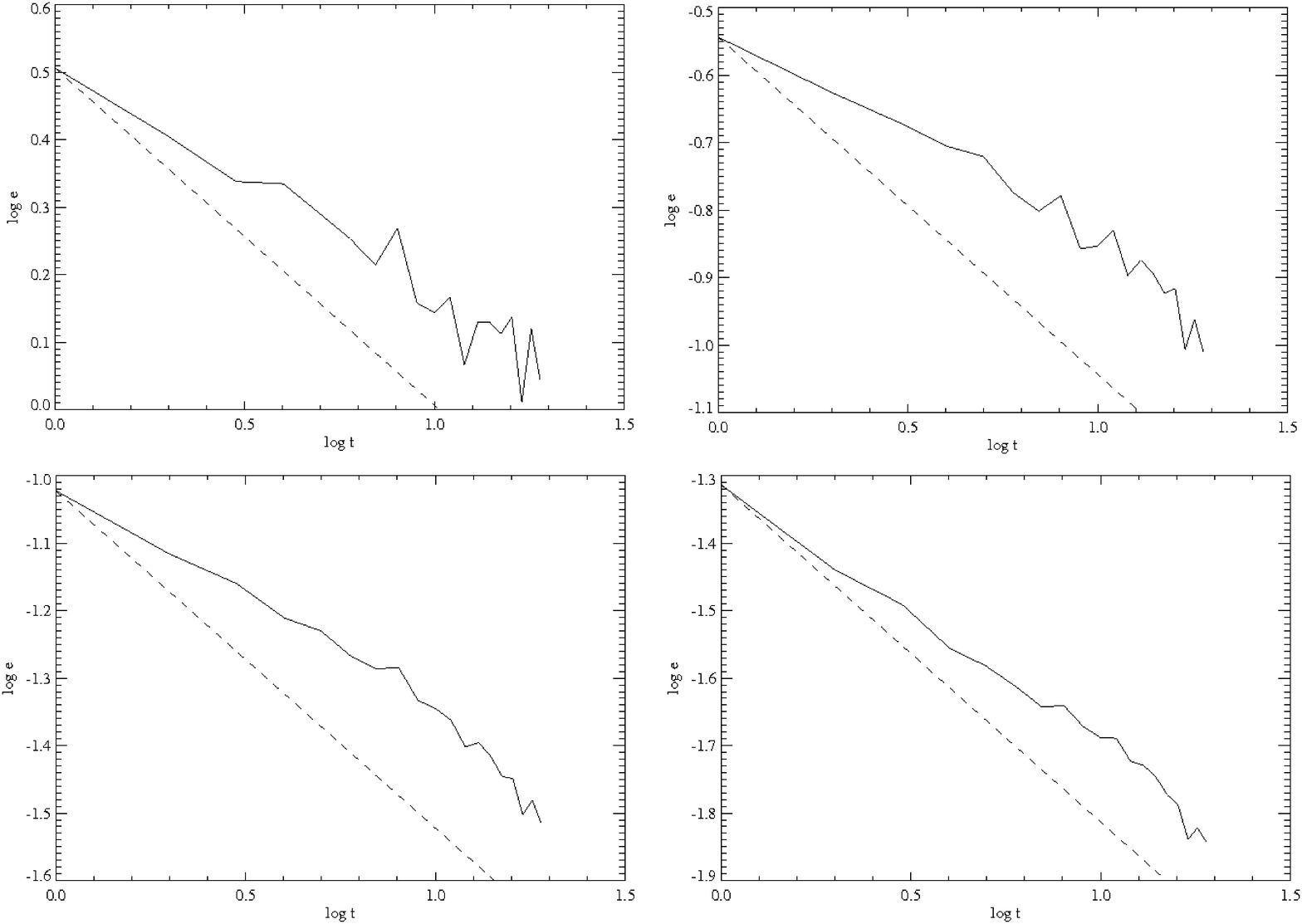}
\caption{The logarithmic Strehl-normalized error $e$ as a function of time for the epoch 2 data (solid line). Upper left: Zone 1 (0-19 pixels away from the center). Upper right: Zone 2 (20-39 pixels). Lower left: Zone 3 (40-59 pixels). Lower right: Zone 4 (60-79 pixels). The dashed line indicates a reference slope corresponding to $e \sim t^{-1/2}$. The errors fall off slower than, but fairly close to, the $e \sim t^{-1/2}$ slope that would be expected for, e.g., photon noise dominated data. Oddly, the errors seem to drop off faster for longer integration times than for shorter times for this particular epoch, whereas the opposite would generally be expected. As would be expected, the dispersion is the largest in the innermost region, where part of the stellar PSF is saturated, and where noise variations are generally larger. Note that the x-axis is in logarithm scale.}
\label{ep2_times}
    \end{figure*}

\begin{figure*}[htb]
   \centering
   \includegraphics[width=16.0cm]{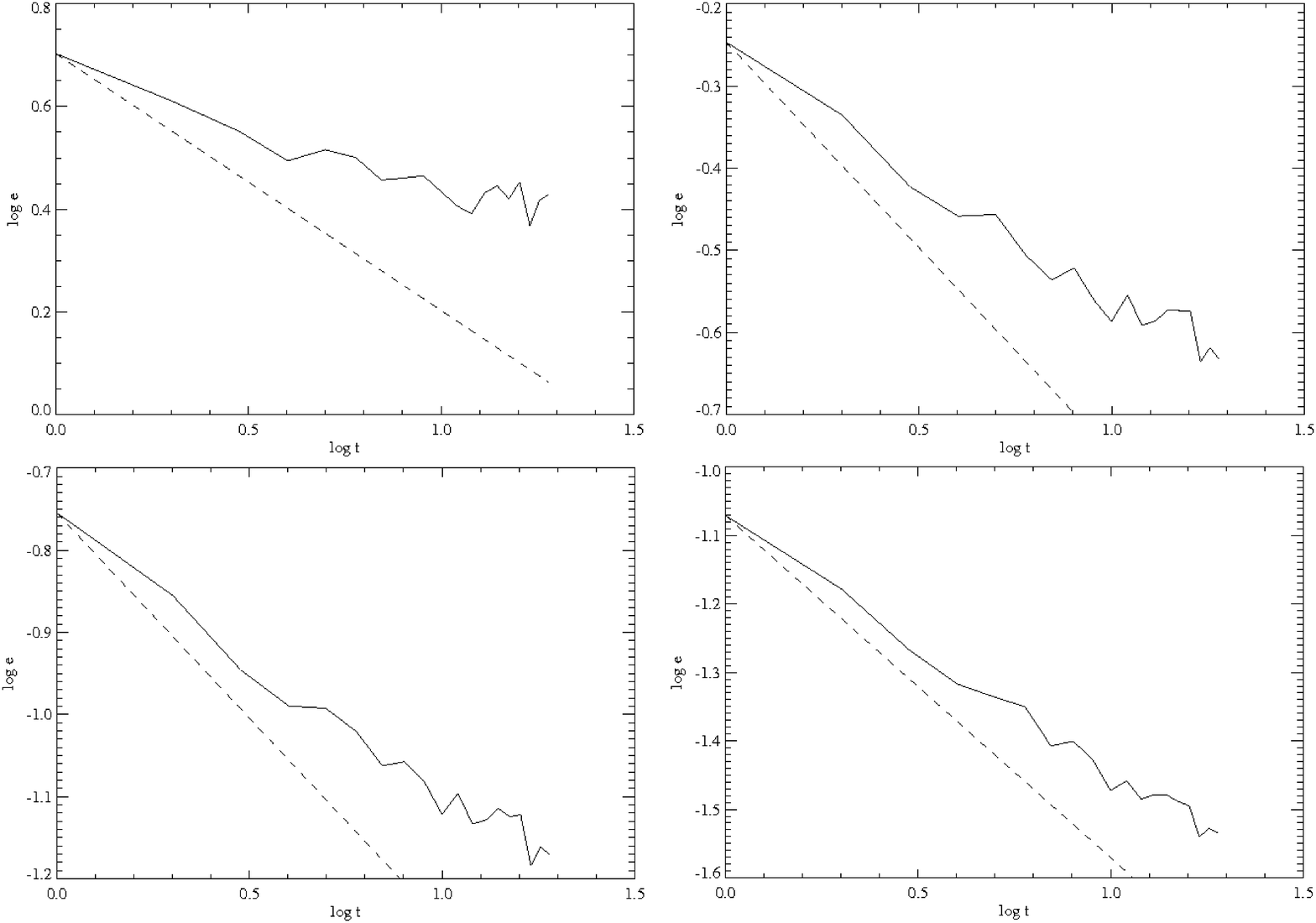}
\caption{The logarithmic Strehl-normalized error $e$ as a function of time for the epoch 3 data (solid line). Upper left: Zone 1 (0-19 pixels away from the center). Upper right: Zone 2 (20-39 pixels). Lower left: Zone 3 (40-59 pixels). Lower right: Zone 4 (60-79 pixels). The dashed line indicates a reference slope corresponding to $e \sim t^{-1/2}$. The errors fall off slower than, but fairly close to, the $e \sim t^{-1/2}$ slope that would be expected for, e.g., photon noise dominated data. As would be expected (and in difference from epoch 2), the error falls off slower at longer integration times, as the residual noise is becoming increasingly dominated by static or quasi-static noise sources. The fall-off seems somewhat better in the outer regions than in the inner ones. Note that the x-axis is in logarithm scale.}
\label{ep3_times}
    \end{figure*}

\begin{figure*}[htb]
   \centering
   \includegraphics[width=16.0cm]{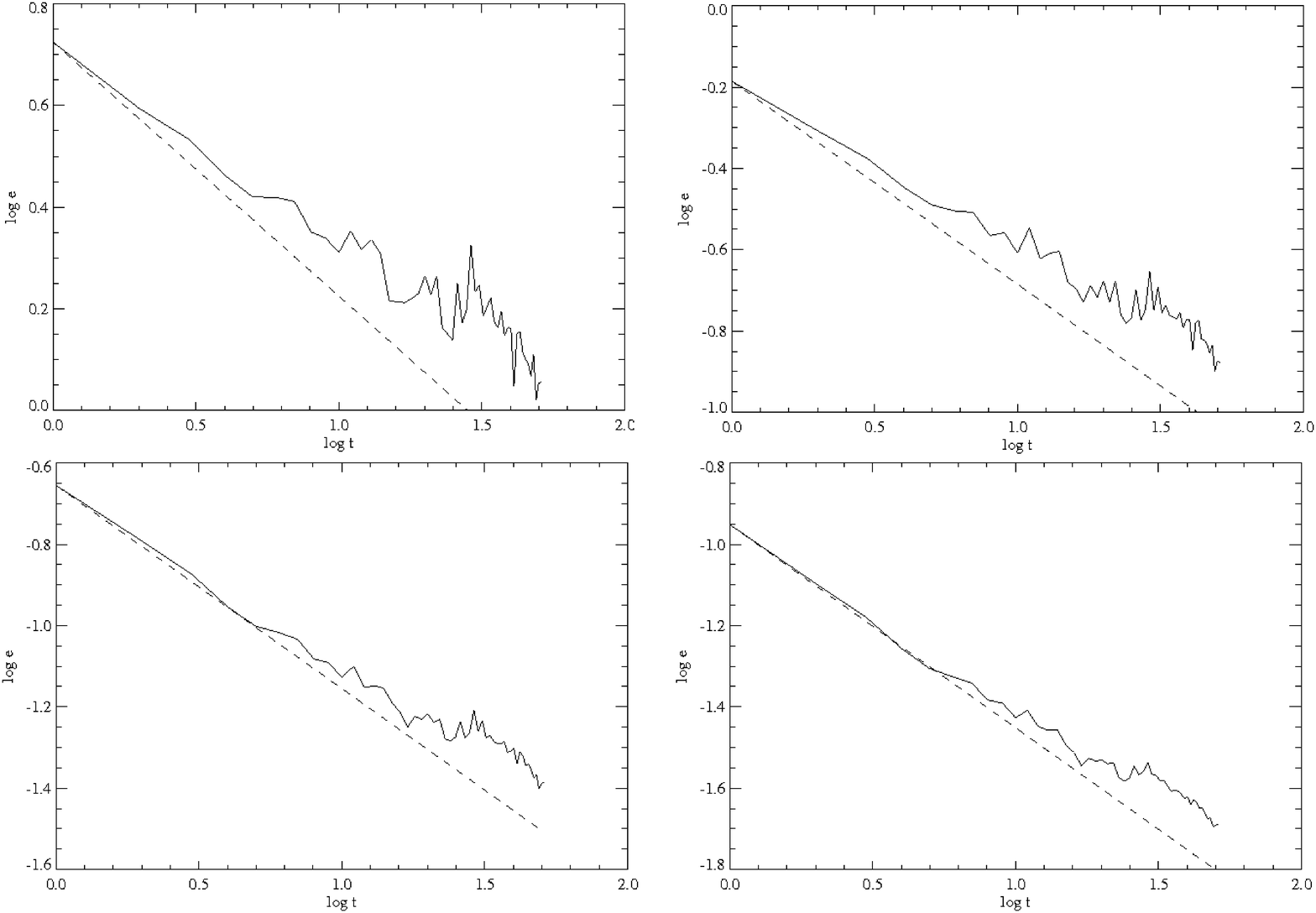}
\caption{The logarithmic Strehl-normalized error $e$ as a function of time for the epoch 4 data (solid line). Upper left: Zone 1 (0-19 pixels away from the center). Upper right: Zone 2 (20-39 pixels). Lower left: Zone 3 (40-59 pixels). Lower right: Zone 4 (60-79 pixels). The dashed line indicates a reference slope corresponding to $e \sim t^{-1/2}$. The errors fall off slower than, but fairly close to, the $e \sim t^{-1/2}$ slope that would be expected for, e.g., photon noise dominated data. Due to the large amount of data points for this epoch, this is the most reliable data set of the three. The fall-off of the errors are slower for larger integration times, as was seen also for epoch 3. The dispersion also clearly decreases outwards from the center. Also, the fall-off seems to be the fastest in the outermost regions. It is remarkable that after 1.5 hours of effective integration time per angle, the error still drops very close to the $e \sim t^{-1/2}$ slope. Note that the x-axis is in logarithm scale.}
\label{ep4_times}
    \end{figure*}

For finding the dependence of $e$ on the effective integration time $t$, each 0$^{\circ}$ frame within one epoch is paired with one 33$^{\circ}$. The pairs are then sorted sequentially in groups of more and more pairs, and the groups are submitted to the full data reduction -- i.e., first groups of one pair per group are formed and reduced, then groups of two pairs per group are formed, then three pairs per group and so on. The average $e$ for a certain number of pairs per group is then plotted against the number of pairs per group. This is shown in Figs. \ref{ep2_times} through \ref{ep4_times}.

To interpret the results of the examination, we need to know the behaviour of the noise sources in the data. As we have mentioned previously, the residual noise in double-differenced (SDI and ADI) data is a mixture of correlated and uncorrelated, dynamic and quasi-static noise. Photon noise and read noise are dynamic and uncorrelated noise sources whose characteristics are well known and easily estimated. They average out with time (for co-added exposures) as $e \sim t^{-1/2}$. Flat field noise is a multiplicative noise source which is static with respect to the detector, but mainly uncorrelated in space. It does not average out with the total number of exposures, but with the amount of different dither positions (five, in our case). It is also completely differenced in the ADI for the cases where the dither positions are the same at 0$^{\circ}$ as at 33$^{\circ}$ (with respect to the center of the PSF). The speckle noise is correlated, and ranges from dynamic to quasi-static. For speckle noise, $\sigma _{\rm s} \sim n_{\rm s}^{-1/2}$, where $\sigma _{\rm s}$ is the standard deviation of the speckle noise and $n_{\rm s}$ is the number of speckles per unit area (see Racine et al. 1999; also, see Sivaramakrishnan et al. 2002 for a specific discussion on time dependence). Thus if the speckle lifetime $\tau _{\rm s}$ is shorter than the integration time of a single exposure ($t _{\rm exp}$), averaging the exposures will yield that $\sigma _{\rm s} \sim t^{-1/2}$, i.e., the component of the speckle noise that varies on such timescales will average out according to the square root law just as for the other dynamic noise sources, because a completely new speckle pattern will be generated in each exposure, such that the number of speckles increases linearly with integration time. Note that it makes no difference that the noise is spatially correlated in this regard -- on a frame-to-frame basis, it will obey Poissonian statistics and average out just like spatially uncorrelated noise. If $\tau _{\rm s} > t _{\rm exp}$, the noise impact will decrease more slowly. In the extreme case where $\tau _{\rm s} > t_{\rm tot}$ (where $t_{\rm tot}$ is the total integration time), the same speckle pattern will be generated in every exposure, hence $n_{\rm s}$ is constant, and thus $\sigma _{\rm s}$ is the same independently of integration time. The noise in the latter case, which represents completely static noise with respect to the observations, should in general be differenced out by SDI and ADI, but noise which is constant in time, but varies in both wavelength and rotation angle of the instrument, could in principle remain in the final data.

\begin{figure}
   \centering
   \includegraphics[width=10.0cm]{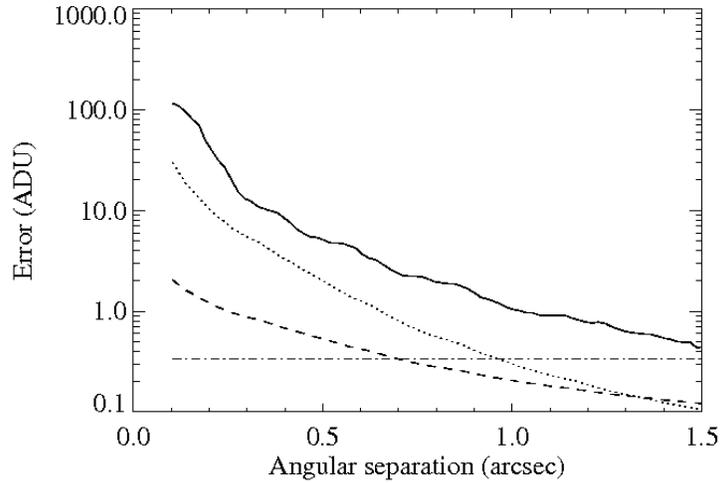}
\caption{The real error (solid line), compared to estimates of the photon noise (dashed line), read noise (dash-dotted line) and the flat field noise (dotted line) for epoch 2. The real noise is dominated by the residual speckle noise, except in the very outermost part where the read noise seems to be significant. The innermost region (within about 0.1 arcsec) is saturated, and does not provide any meaningful information about the real error.}
\label{ep2_theo_errs}
    \end{figure}

\begin{figure*}[htb]
   \centering
   \includegraphics[width=10.0cm]{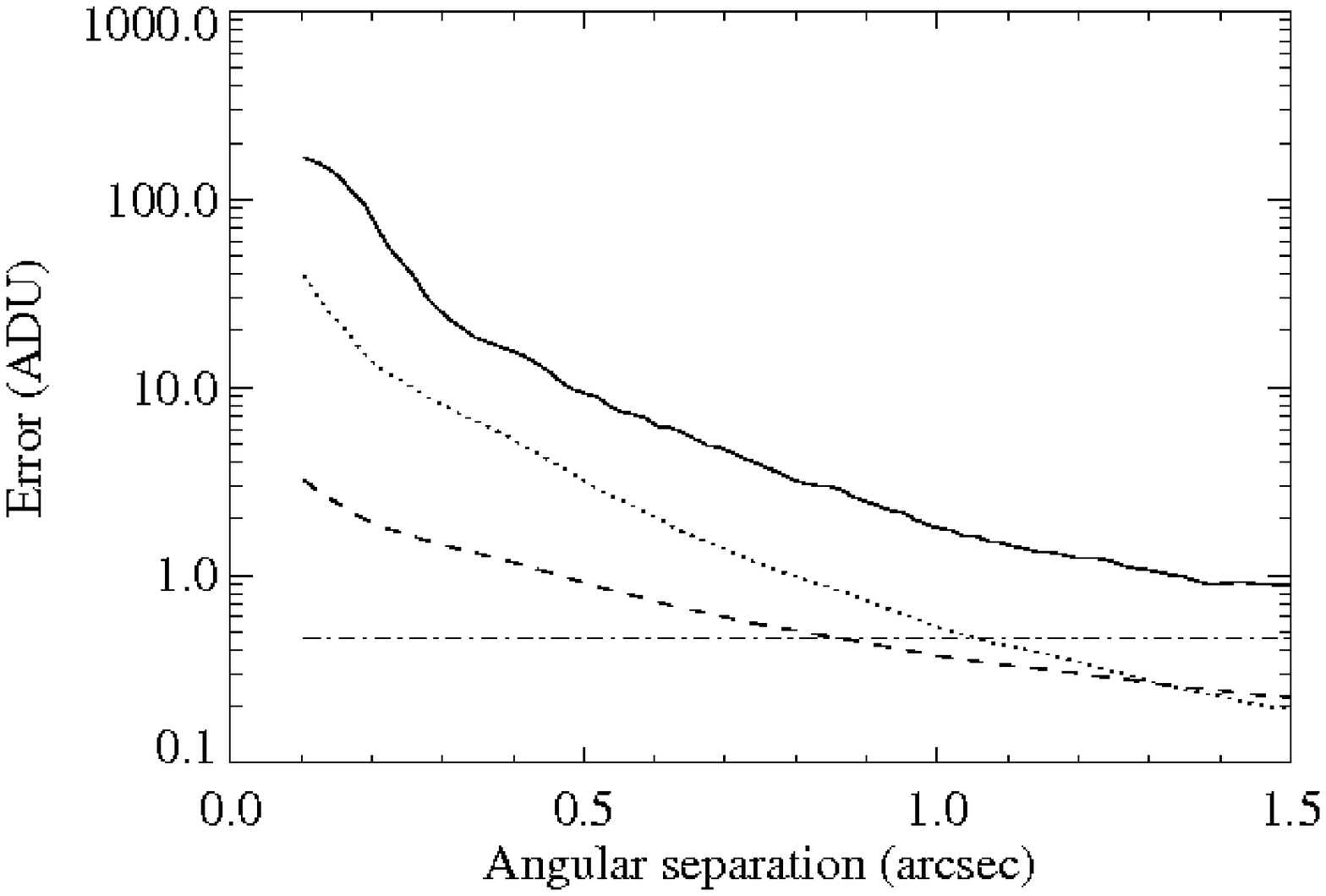}
\caption{The real error (solid line), compared to estimates of the photon noise (dashed line), read noise (dash-dotted line) and the flat field noise (dotted line) for epoch 3. The real noise is dominated by the residual speckle noise. The innermost region (within about 0.1 arcsec) is saturated, and does not provide any meaningful information about the real error.}
\label{ep3_theo_errs}
    \end{figure*}

\begin{figure*}[htb]
   \centering
   \includegraphics[width=10.0cm]{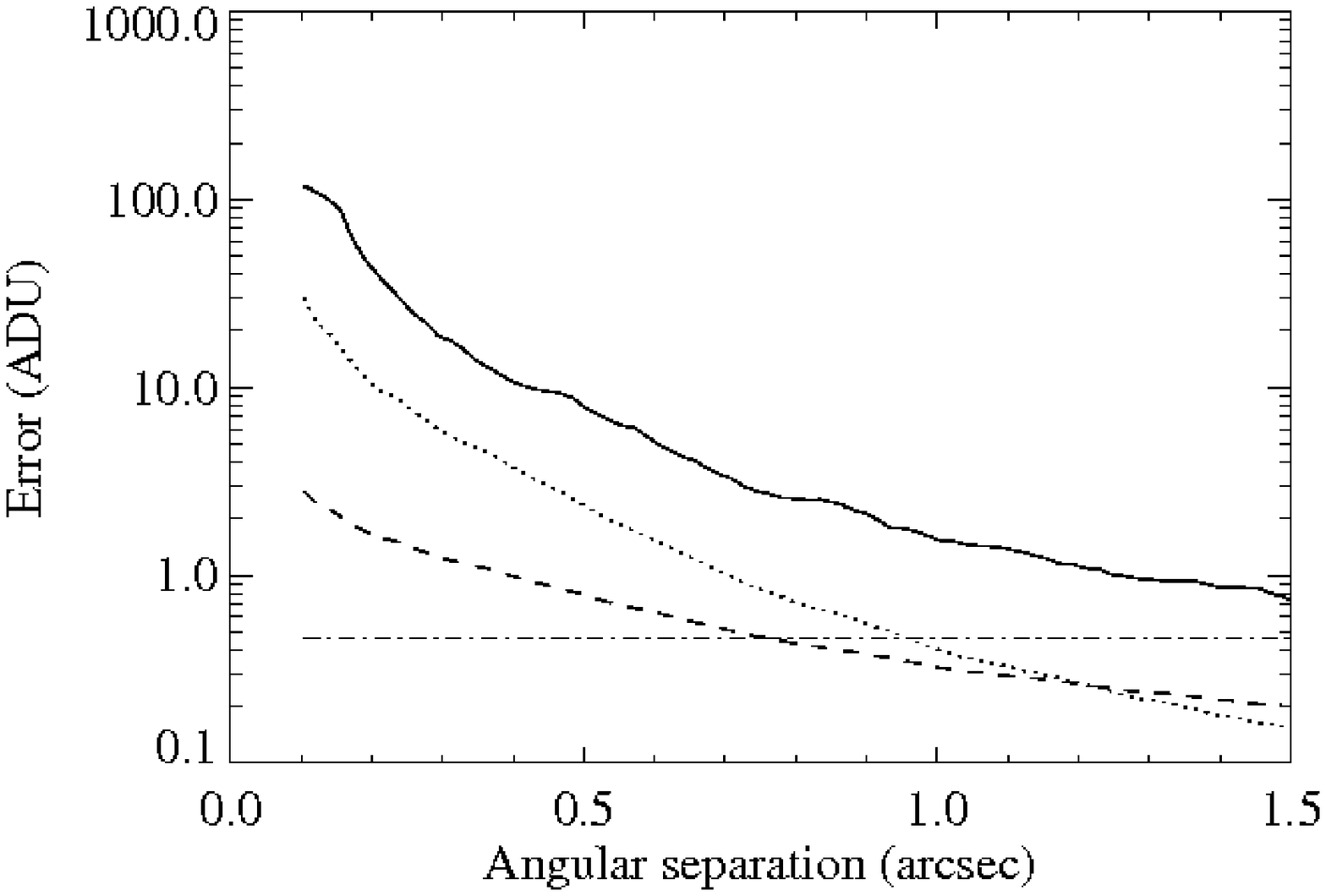}
\caption{The real error (solid line), compared to estimates of the photon noise (dashed line), read noise (dash-dotted line) and the flat field noise (dotted line) for epoch 4. The real noise is dominated by the residual speckle noise. The innermost region (within about 0.1 arcsec) is saturated, and does not provide any meaningful information about the real error.}
\label{ep4_theo_errs}
    \end{figure*}

It is clear from the images of $\epsilon$ Eri (Fig. \ref{ee_epochs}) that the total noise is dominated by correlated noise for most parts of the observed parameter space. This can also be seen in Figs \ref{ep2_theo_errs} through \ref{ep4_theo_errs}, where we have computed the expected photon noise, flat field noise and read noise for each observation, and plotted along with the actual noise for epochs 2, 3, and 4. We see that indeed, the uncorrelated noise is dominated by other noise sources. The photon noise and read noise are strongly dominated in the inner parts, and start to become significant only in the outer parts. The flat field noise is dominated by about the same factor everywhere, which implies that the dominating noise has the same flux dependence as flat field noise -- i.e., that the dominating noise is linearly proportional to the local flux, which indeed is the case for speckle noise (see Racine et al. 1999 and Aime \& Soummer 2004 for halo-dominated images).

Returning to the analysis of the noise trend as a function of time in our real data (Figs \ref{ep2_times} through \ref{ep4_times}), we see that the general trend is a drop which is slightly slower than $e \sim t^{-1/2}$. This is considerably better than expected, and implies that a large fraction of the residual speckle noise has a short lifetime. Obviously, since the speckle noise with the shortest lifetimes will cancel out faster than the more long-lived components, the residual noise will gradually be more and more dominated by quasi-static noise until a noise floor is hit and no further improvement can be gained in terms of integration time. Judging from the curves, that point is however still quite far off. It is particularly interesting that $e$ still drops off close to the $t^{-1/2}$ rate in the epoch 4 data, after about 1.5 hours effective integration time per angle. This implies that a yet higher sensitivity can be reached by simply integrating for a longer time, as long as the other observing conditions are acceptable.

In reference to future strategies related to NACO-SDI imaging, it is clear that a large amount of integration time is favourable. In particular, if the position of the suspected companion is known a priori, as will henceforth be the case for $\epsilon$ Eri b, this is best done by increasing the integration time of individual exposures (DIT), since this minimizes the readout time, and is more efficient for mitigation of read noise. In the general case, where no such a priori information is available, the DIT is always conservatively set such that the primary PSF will ony saturate slightly, in order to maintain an as small inner working angle as possible. However, if the separation is known, the DIT can be set such that the primary PSF saturates over a large area, as long as this area is well within the expected separation. Of course, in such a case, considerations should also be taken about whether there may be additional interesting companions within the separation of the known companion.

\section{Conclusions}

\label{sec_conclusions}

We have performed a multi-epoch study of $\epsilon$ Eri with NACO-SDI at the VLT, and combined it with astrometry in order to try to detect its planetary companion, $\epsilon$ Eri b. Despite excellent H-band contrasts of 14.5 to 15.1 mag at the expected positions of $\epsilon$ Eri b, and limiting absolute magnitudes of 19.1 to 19.5 mag, we did not detect the companion. A theoretical assessment of the brightness based on the mass of $\epsilon$ Eri b, and the plausible age range of the $\epsilon$ Eri system, indicated that a non-detection might perhaps indeed be expected, though such an analysis is necessarily vastly uncertain. With a more well-constrained astrometry, even better detection limits may be possible to achieve from the existing data, through speckle number-count statistics over all four epochs.

In addition, the detection limits as a function of Strehl ratio and integration time have been examined. It has been found that the signal-to-noise ratio scales linearly with the Strehl ratio, which shows that it is of significant importance to maintain a high Strehl ratio during companion searches. A surprising result was reached in the case of detection limit dependence on integration time. The signal-to-noise ratio was found to scale almost according to the well-known square-root dependence for standard noise sources. This means that it may be possible to detect much fainter objects by simply increasing the integration time for a given target. Consequentially, we conclude that with a sufficient amount of effort, objects like $\epsilon$ Eri b may be detectable with the presently available telescopes and instrumentation.

\acknowledgments

Support was provided by the Deutsches Zentrum f\"ur Luft- und Raumfahrt (DLR), F\"orderkennzeichen 50 OR 0401. We wish to thank Andrea Stolte for useful discussion. Markus Janson gratefully receives financial support from the International Max Planck Research School (IMPRS), Heidelberg.




\clearpage




\clearpage

\clearpage

\clearpage

\clearpage

\clearpage

\clearpage

\clearpage

\clearpage

\clearpage

\clearpage

\clearpage

\clearpage

\clearpage

\begin{deluxetable}{lllllllll}
\tabletypesize{\scriptsize}
\rotate
\tablecaption{Observing log of NACO high-contrast imaging observations of $\epsilon$ Eri \label{tab1}}
\tablewidth{0pt}
\tablehead{
\colhead{Epoch} & \colhead{Main date} & \colhead{MJD} & \colhead{Frames per angle} & \colhead{DIT(s)} & \colhead{NDIT} & \colhead{Tot. time per angle} & \colhead{Mean Strehl (1.6 $\mu$m)} & \colhead{Mean seeing}
}
\startdata
1 & 17 Aug. 2003 & 52868 & 10 & 0.5 & 60 & 300 & 11.5\% & 1.26 \\
2 & 19 Sep. 2004 & 53267 & 20 & 0.6 & 160 & 1920 & 32.2\% & 0.91 \\
3 & 10 Aug. 2005 & 53592 & 16 & 1.0 & 86 & 1376 & 35.7\% & 0.87 \\
4 & 1 Jan. 2006 & 53736 & 52 & 1.0 & 86 & 4472 & 33.8\% & 0.82 \\
\enddata
\end{deluxetable}

\clearpage

\begin{deluxetable}{llll}
\tabletypesize{\scriptsize}
\tablecaption{Expected separation and position angle for each epoch from the astrometry of Benedict et al. (2006). \label{tab3}}
\tablewidth{0pt}
\tablehead{
\colhead{Epoch} & \colhead{Date} & \colhead{Sep. (arcsec)} & \colhead{P.A. (deg)}
}
\startdata
1 & 17 Aug. 2003 & $1.68 \pm 0.18$ & $114 \pm 8$ \\
2 & 19 Sep. 2004 & $1.55 \pm 0.18$ & $129 \pm 8$ \\
3 & 10 Aug. 2005 & $1.27 \pm 0.15$ & $144 \pm 8$ \\
4 & 1 Jan. 2006 & $1.09 \pm 0.13$ & $155 \pm 9$ \\
\enddata
\end{deluxetable}

\clearpage

\begin{deluxetable}{llll}
\tabletypesize{\scriptsize}
\tablecaption{Expected H-band brightness contrast between $\epsilon$ Eri A and b for different ages, from the Baraffe et al. (2003) models. \label{tab2}}
\tablewidth{0pt}
\tablehead{
\colhead{Age} & \colhead{Contrast} & \colhead{Planet temp.} & \colhead{Planet radius}
}
\startdata
10 Myr & 12.5 mag & 640 K & 0.135 $R_{\rm sun}$ \\
50 Myr & 15.5 mag & 430 K & 0.123 $R_{\rm sun}$ \\
120 Myr & 17.4 mag & 350 K & 0.118 $R_{\rm sun}$ \\
500 Myr & 22.9 mag & 240 K & 0.111 $R_{\rm sun}$ \\
1 Gyr & 26.0 mag & 190 K & 0.108 $R_{\rm sun}$ \\
\enddata
\end{deluxetable}

\end{document}